\documentclass[twocolumn,showpacs,preprintnumbers,amsmath,amssymb]{revtex4}
\usepackage{graphicx}
\usepackage{dcolumn}
\usepackage{bm}

\newcommand{\ha}{\hat{a}}
\newcommand{\hA}{\hat{A}}
\newcommand{\hAD}{\hat{\cal A}}

\newcommand{\hap}{\hat{a}^+}
\newcommand{\hAp}{\hat{A}^+}
\newcommand{\hApD}{\hat{\cal A}^+}

\newcommand{\hB}{\hat{\cal B}}
\newcommand{\hBp}{\hat{\cal B}^+}

\newcommand{\hC}{\hat{\cal C}}
\newcommand{\hCp}{\hat{\cal C}^+}

\newcommand{\hH}{\hat{H}}
\newcommand{\hHD}{\hat{\cal H}}

\newcommand{\hO}{\hat{\Omega}}
\newcommand{\hM}{\hat{\cal M}}

\newcommand{\hY}{\hat{\cal Y}}
\newcommand{\hX}{\hat{\cal X}}

\begin{document}
\title{Zitterbewegung of relativistic electrons in a magnetic field and its simulation by trapped ions}
\date{\today}
\author{Tomasz M. Rusin $^{1,3}$}
\email{Tomasz.Rusin@centertel.pl}
\author{Wlodek Zawadzki$^2$}
\affiliation{$^1$ PTK Centertel sp. z o.o., ul. Skierniewicka 10A, 01-230 Warsaw, Poland\\
             $^2$ Institute of Physics, Polish Academy of Sciences, Al. Lotnik\'ow 32/46, 02-688 Warsaw, Poland \\
             $^3$ Orange Customer Service sp. z o. o., ul. Twarda 18, 00-105 Warsaw, Poland}

\begin{abstract}
One-electron 3+1 and 2+1 Dirac equations are used to calculate the motion
of a relativistic electron in a vacuum in the presence of an external
magnetic field. First, calculations are carried on an operator level and
exact analytical results are obtained for the electron trajectories which
contain both intraband frequency components, identified as the cyclotron
motion, as well as interband frequency components, identified as the
trembling motion (Zitterbewegung, ZB). Next, time-dependent Heisenberg
operators are used for the same problem to compute average values of
electron position and velocity employing Gaussian wave packets. It is
shown that the presence of a magnetic field and the resulting quantization
of the energy spectrum has pronounced effects on the electron
Zitterbewegung: it introduces intraband frequency components into the
motion, influences all the frequencies and makes the motion stationary
(not decaying in time) in case of the 2+1 Dirac equation. Finally,
simulations of the 2+1 Dirac equation and the resulting electron ZB in
the presence of a magnetic field are proposed and described employing
trapped ions and laser excitations. Using simulation parameters achieved
in recent experiments of Gerritsma and coworkers we show that the effects
of the simulated magnetic field on ZB are considerable and can certainly
be observed.
\end{abstract}

\maketitle
\section{Introduction}

The phenomenon of Zitterbewegung~(ZB) for free relativistic electrons in a vacuum goes back
to the work of Schrodinger,
who showed in 1930 that, due to a non-commutativity of the velocity operators with the Dirac Hamiltonian,
relativistic electrons experience a trembling motion in absence of external fields~\cite{Schroedinger1930}.
The ZB is a strictly quantum phenomenon as it goes beyond Newton's first
law of classical motion. Since the Schrodinger prediction the subject of ZB was treated by very many
theoretical papers. It was recognized that ZB is due to an interference of electron states with positive
and negative electron energies~\cite{BjorkenBook,ThallerBook}.
The frequency of ZB oscillations predicted by Schrodinger is very high,
corresponding to~$\hbar\omega_Z \simeq 2mc^2$, and its amplitude is very small, being around the Compton
wavelength~$\hbar/mc = 3.86\times$ 10$^{-3}$~\AA.
Thus, it is impossible to observe this effect in its original form with the currently available experimental means.
In fact, even the principal observability of ZB in a vacuum was often questioned in the literature
\cite{Huang1952,Krekora2004}.
However, in a very recent paper Gerritsma {\it et al.}~\cite{Gerritsma2010} simulated the 1+1 Dirac equation
(DE) and the resulting Zitterbewegung with the use of trapped Ca ions excited by
appropriate laser beams. The remarkable advantage of this method is that one can simulate the
basic parameters of DE, i.e.~$mc^2$ and~$c$, and give them desired values. This results in a much lower
ZB frequency and a much larger ZB amplitude. The simulated values were in fact experimentally observed.

The general purpose of our work is concerned with the electron ZB in the presence of a magnetic field.
The presence of a constant magnetic field
does not cause electron transitions between negative and positive electron energies. On the other hand,
it quantizes the energy spectrum into Landau levels which brings qualitatively new features into the ZB.
Our work has three objectives. First, we calculate the Zitterbewegung of relativistic electrons in a vacuum in the
presence of an external magnetic field at the operator level. We obtain exact analytical formulas for this problem.
Second, we calculate average values describing ZB of an electron prepared in the form of a Gaussian wave packet.
These average values can be directly
related to possible observations. However, as mentioned above and confirmed by our calculations, the corresponding
frequencies and amplitudes of ZB in a vacuum are not accessible experimentally at present.
For this reason, and this is our third objective,
we propose and describe simulations of ZB in the presence of a magnetic field with the use of trapped ions.
We do this keeping in mind the recent experiments reported by Gerritsma~{\it et al.}. We show that,
employing the simulation parameters
of Ref.~\cite{Gerritsma2010}, one should be able to observe the magnetic effects in ZB.

The problem of ZB in a magnetic field was treated
before~\cite{Barut1985}, but the results were limited to the operator level and suffered
from various deficiencies which we mention in Appendix~\ref{AppendixBarut}.
A similar problem was treated in Ref.~\cite{Villavicencio2000} at the operator level
in weak magnetic field limit. Bermudez {\it et al.}~\cite{Bermudez2007}
treated a related problem of mesoscopic superposition states in relativistic Landau levels.
We come back to this work in the Discussion.
Our treatment aims to calculate directly the observable Zitterbewegung effects.
Preliminary results of our work were published in Ref.~\cite{Rusin2010}.

An important aspect of ZB, which was not considered in the pioneering work of Schrodinger and most of the
papers that followed it, is an existence of the 'Fermi sea' of electrons filling negative energy states.
This feature can seriously affect the phenomenon of ZB, see~\cite{Krekora2004}.
We emphasize that both our calculations
as well as the simulations using trapped ions~\cite{Gerritsma2010} are based
on the 'empty Dirac equation' for which ZB certainly exists. We come back to this problem in the Discussion.

Our paper is organized in the following way. In Section II we use the 3+1 Dirac equation
to derive the time dependence of operators describing
motion of relativistic electrons in a vacuum in the presence of a
magnetic field. Intraband frequency components (the cyclotron motion)
are distinguished from interband frequency components (the trembling
motion). In Sections III and IV we treat the same subject calculating averages
of the time-dependent Heisenberg operators with the use of Gaussian wave
packets. This formulation is more closely related to possible
experiments. In Section V we simulate the 2+1 Dirac equation and the
resulting electron Zitterbewegung employing trapped ions and laser
excitations in connection with the recent experimental simulation of
electron ZB in absence of magnetic field. In Section VI we discuss our
results. The paper is concluded by a summary. In appendices we discuss
some technical aspects of the calculations and the relation
of our work to that of other authors.

\section{Zitterbewegung: Operator form}
We consider a relativistic electron in a magnetic field. Its Hamiltonian is
\begin{equation} \label{H_DE}
\hHD = c\alpha_x\hat{\pi}_x + c\alpha_y\hat{\pi}_y + c\alpha_z\hat{\pi}_z + \beta mc^2,
\end{equation}
where ${\bm \hat{\pi}} = {\bm \hat{p}}- q{\bm A}$ is the generalized momentum,~$q$
is the electron charge,~$\alpha_i$ and~$\beta$ are Dirac matrices in the standard notation.
Taking the magnetic field ${\bm B} \| {\bm z}$ we choose the vector potential ${\bm A} = (-By,0,0)$.
For an electron there is~$q=-e$ with~$e>0$. One can look for solutions in the form
\begin{equation} \Psi({\bm r}) = e^{ik_xx+ik_zz}\Phi(y), \end{equation}
and we obtain an effective Hamiltonian~$\hHD$
\begin{equation}
 \hHD = c\hbar\left[(k_x - eBy/\hbar)\alpha_x + (\partial/i\partial y) \alpha_y + k_z\alpha_z \right] + \beta mc^2.
\end{equation}
Introducing the magnetic radius $L = \sqrt{\hbar/eB}$ and $\xi=y/L-k_xL$ we have
$y=\xi L + k_xL^2$, $eB/\hbar=1/L^2$,
and $\partial/ \partial y = (1/L)\partial/ \partial \xi$.
Defining the standard raising and lowering operators for the harmonic oscillator
\begin{equation}
 \left\{\begin{array}{ccc} \label{H_aap_def}
     \ha  &=& (\xi+ \partial/\partial \xi)/\sqrt{2},  \\
     \hap &=& (\xi -\partial/\partial \xi)/\sqrt{2},   \end{array}\right.
\end{equation}
one has $[\ha,\hap]=1$ and $\xi = (\ha + \hap)/\sqrt{2}$. The Hamiltonian~$\hHD$ reads
\begin{equation}
\hHD = \left(\begin{array}{cc}  mc^2\hat{\bm 1}   & \hH + E_z\sigma_z  \\
    \hH + E_z\sigma_z & -mc^2 \hat{\bm 1} \\  \end{array} \right),
\end{equation}
where~$\hat{\bm 1}$ is the 2$\times$2 identity matrix, $E_z=c\hbar k_z$, and
\begin{equation} \label{H aap}
 \hH = -\hbar\omega\left(\begin{array}{cc}  0 & \ha \\ \hap & 0 \\  \end{array}\right),
\end{equation}
with~$\omega=\sqrt{2}c/L$. The frequency~$\omega$ (which should not be confused with the cyclotron
frequency~$\omega_c=eB/m$) is often used in our considerations.

Now we introduce an important four-component operator
\begin{equation} \label{H_hAD_def} \hAD = {\rm diag}(\hA,\hA), \end{equation}
where $\hA = {\rm diag}(\ha,\ha)$. Its adjoint operator is
\begin{equation} \label{H_hApD_def} \hApD = {\rm diag} (\hAp,\hAp), \end{equation}
where $\hAp ={\rm diag}(\hap,\hap)$.
Next we define the four-component position operators
\begin{eqnarray}
  \hY &=& \frac{L}{\sqrt{2}} \left(\hAD + \hApD\right), \label{H_Y}\\
  \hX &=& \frac{L}{i\sqrt{2}}\left(\hAD - \hApD\right), \label{H_X}
\end{eqnarray}
in analogy to the position operators~$\hat{y}$ and~$\hat{x}$, see Appendix~\ref{AppendixXY}.
We intend to calculate
the time dependence of~$\hAD$ and~$\hApD$ and then the time dependence of~$\hY$ and~$\hX$.

To find the dynamics of~$\hAD$ we calculate the first and second time derivatives of~$\hAD$ using the
equation of motion: $\hAD_t\equiv d\hAD/dt = (i/\hbar)[\hHD,\hAD]$.
Since~$\hat{\bm 1}$ and~$\sigma_z$ commute with~$\hA$ and~$\hAp$, we obtain
\begin{eqnarray}
 \hAD_t &=& \frac{i}{\hbar} \left(\begin{array}{cc} 0 & [\hH,\hA] \\ \protect{[\hH,\hA]} & 0  \end{array} \right),
 \\
 \hApD_t &=& \frac{i}{\hbar} \left(\begin{array}{cc} 0 & [\hH,\hAp] \\ \protect{[\hH,\hAp]} & 0  \end{array} \right).
\end{eqnarray}
There is $(i/\hbar)[\hH,\hA] = \hA_t = i\omega \left(\begin{array}{cc} 0 & 0\\  1 & 0 \end{array}\right)$
and $\hAp_t = -i\omega \left(\begin{array}{rr} 0 & 1\\  0 & 0 \end{array}\right)$.
In consequence
\begin{eqnarray}
 \hAD_t  &=& \left(\begin{array}{cc} 0 & \hA_t  \\ \hA_t  & 0  \end{array} \right),   \\
 \hApD_t &=& \left(\begin{array}{cc} 0 & \hAp_t \\ \hAp_t & 0  \end{array} \right).
\end{eqnarray}

The second time derivatives of~$\hAD$ and~$\hApD$ are calculated following the trick proposed by Schrodinger.
We use two versions of this trick
\begin{eqnarray}
\hAD_{tt} =&(i/\hbar) [\hHD, \hAD_t] =&
   \frac{2i}{\hbar} \hHD \hAD_t - \frac{i}{\hbar}\{\hHD,\hAD_t \}, \\
\hApD_{tt} =& (i/\hbar) [\hHD, \hApD_t] =&
   -\frac{2i}{\hbar} \hApD_t\hHD + \frac{i}{\hbar}\{\hHD, \hApD_t\}.
\end{eqnarray}
The anticommutator of~$\hAD_t$ and~$\hHD$ is
\begin{eqnarray}
\frac{i}{\hbar}\{\hHD,\hAD_t \}
  &=&  \frac{i}{\hbar}
   \left(\begin{array}{cc} \{\hA_t,\hH\} & 0  \\  0  & \{\hA_t,\hH\}  \end{array} \right).
\end{eqnarray}
Similarly
\begin{eqnarray}
\frac{i}{\hbar}\{\hHD,\hApD_t \}   &=& \frac{i}{\hbar}
  \left(\begin{array}{cc} \{\hAp_t,\hH\} & 0  \\  0  & \{\hAp_t,\hH\}  \end{array} \right).
\end{eqnarray}
We need to know the anticommutators $\{\hH,\hA_t\}$ and $\{\hH,\hAp_t\}$. There is
$(i/\hbar) \{\hH,\hA_t \}  =  \omega^2 \hA$ and
$(i/\hbar) \{\hH,\hAp_t \} = -\omega^2 \hAp$, so that
\begin{eqnarray} \label{H_ac1}
-\frac{i}{\hbar}\{\hHD,\hAD_t \}   &=& \omega^2  \left(\begin{array}{cc} \hA & 0  \\  0  & \hA  \end{array} \right),
 \\              \label{H_ac2}
\frac{i}{\hbar}\{\hHD,\hApD_t \}  &=& -\omega^2  \left(\begin{array}{cc} \hAp & 0  \\  0  & \hAp  \end{array} \right).
\end{eqnarray}
Thus we finally obtain from Eqs.~(\ref{H_ac1}) and~(\ref{H_ac2}) second order equations for~$\hAD$ and~$\hApD$
\begin{eqnarray}
  \hAD_{tt} &=&  (2i/\hbar) \hHD\hAD_t  - \omega^2 \hAD  \\
  \hApD_{tt}&=& -(2i/\hbar) \hApD_t\hHD - \hApD\omega^2 .
\end{eqnarray}
To solve the above equations we eliminate the terms with the first derivative using the substitutions
$\hAD = \exp(+i\hHD t/\hbar)\hB$ and $\hApD = \hBp\exp(-i\hHD t/\hbar)$, which gives
\begin{eqnarray}
   \hB_{tt}    &=& -(1/\hbar^2)\hHD^2\hB - \omega^2\hB,. \\
   \hBp_{tt}   &=& -(1/\hbar^2)\hBp\hHD^2 - \hBp\omega^2.
\end{eqnarray}
Finally
\begin{eqnarray}
 \label{H_Btt}  \hB_{tt}  &=& -(\hO^2 + \omega^2)\hB,  \\
 \label{H_Bptt} \hBp_{tt} &=& -\hBp (\hO^2 + \omega^2),
\end{eqnarray}
where~$\hO = \hHD/\hbar$. The solutions of the above equations are
\begin{eqnarray}
 \hB  &=&        e^{-i\hM t}\hC_1 +   e^{i\hM t}\hC_2, \\
 \hBp &=& \hCp_1e^{-i\hM t} +  \hCp_2e^{i\hM t},
\end{eqnarray}
where $\hM =+\sqrt{\hO^2+\omega^2}$ is the positive root of $\hM^2 =\hO^2+\omega^2$.
The operator~$\hM$ is an important quantity in our
considerations. Both~$\hC_1$ and~$\hCp_2$ are time-independent operators.
Coming back to~$\hAD(t)$ and~$\hApD(t)$ we have
\begin{eqnarray}
   \hAD(t) &=& e^{i\hO t}e^{-i\hM t}\hC_1+ e^{i\hO t}e^{+i\hM t}C_2,  \\
  \hApD(t) &=& \hCp_1 e^{+i\hM t}e^{-i\hO t}+\hCp_2 e^{-i\hM t}e^{-i\hO t}.
\end{eqnarray}
In order to find the final
forms of~$\hAD(t)$ and~$\hApD(t)$ one has to use the initial conditions. They are
\begin{eqnarray}
\hAD(0) &=& \hC_1 + \hC_2,                               \nonumber \\
\hApD(0) &=& \hCp_1 + \hCp_2,                            \nonumber \\
\hAD_t(0) &=& i(\hO - \hM) \hC_1+ i(\hO + \hM)\hC_2,     \nonumber \\
\hApD_t(0) &=& -i\hCp_1(\hO - \hM)  -i\hCp_2(\hO + \hM). \nonumber
\end{eqnarray}
Simple manipulations give
\begin{eqnarray} \label{H_Init_hC}
  \hC_1 &=&  \frac{i}{2}\hM^{-1}\hAD_t(0) + \frac{1}{2}\hM^{-1}\hO\hAD(0)+\frac{1}{2}\hAD(0),\\
  \hC_2 &=& -\frac{i}{2}\hM^{-1}\hAD_t(0) - \frac{1}{2}\hM^{-1}\hO\hAD(0)+\frac{1}{2}\hAD(0).
\end{eqnarray}
Similarly
\begin{eqnarray}
 \label{H_Init_hCp}
  \hCp_1 &=&  -\frac{i}{2}\hApD_t(0)\hM^{-1} + \frac{1}{2}\hApD(0)\hO\hM^{-1}+\frac{1}{2}\hApD(0),\ \ \ \  \\
  \hCp_2 &=&   \frac{i}{2}\hApD_t(0)\hM^{-1} - \frac{1}{2}\hApD(0)\hO\hM^{-1}+\frac{1}{2}\hApD(0).\ \ \ \
\end{eqnarray}
One can see by inspection that the initial conditions for~$\hAD(0)$ and~$\hAD_t(0)$ are satisfied.
It is convenient to express~$\hAD_t$ in terms of~$\hAD$ and~$\hO$ using the equation
of motion $i\hAD_t=\hAD\hO - \hO\hAD$.
Then the first and second terms in Eqs.~(\ref{H_Init_hC}) and~(\ref{H_Init_hCp}) partially cancel out
and the operator~$\hAD(t)$ can be expressed as a sum $\hAD(t)=\hAD_1(t)+\hAD_2(t)$, where
\begin{eqnarray}
  \label{H_Init_A1} \hAD_1(t) &=&  \frac{1}{2}e^{i\hO t}e^{-i\hM t} \left[\hAD(0) + \hM^{-1}\hAD(0)\hO \right], \\
  \label{H_Init_A2} \hAD_2(t) &=&  \frac{1}{2}e^{i\hO t}e^{+i\hM t} \left[\hAD(0) - \hM^{-1}\hAD(0)\hO\right].
\end{eqnarray}
Similarly, one can break $\hApD(t)=\hApD_1(t)+\hApD_2(t)$, where
\begin{eqnarray}
  \label{H_Init_Ap1} \hApD_1(t) &=&  \frac{1}{2}\left[\hApD(0) + \hO\hApD(0)M^{-1}\right]e^{+i\hM t}e^{-i\hO t},\ \ \\
  \label{H_Init_Ap2} \hApD_2(t) &=&  \frac{1}{2}\left[\hApD(0) - \hO\hApD(0)M^{-1}\right]e^{-i\hM t}e^{-i\hO t}.\ \
\end{eqnarray}
Using Eqs.~(\ref{H_Y}) and~(\ref{H_X}) we obtain
\begin{eqnarray}
  \hY(t)&=& \frac{L}{\sqrt{2}} \left(\hAD_1(t) + \hAD_2(t) + \hApD_1(t) + \hApD_2(t)\right), \label{H_Yt} \ \ \\
  \hX(t)&=& \frac{L}{i\sqrt{2}}\left(\hAD_1(t) + \hAD_2(t) - \hApD_1(t) - \hApD_2(t)\right). \label{H_Xt} \ \
\end{eqnarray}

The above compact equations are our final expressions for the time dependence of~$\hAD(t)$ and~$\hApD(t)$
operators and, by means of Eqs.~(\ref{H_Yt}) and~(\ref{H_Xt}), for the time dependence of the position
operators~$\hY(t)$ and~$\hX(t)$. These equations are exact and, as such, they are
quite fundamental for relativistic electrons in a magnetic field.
The results are given in terms of operators~$\hO$ and~$\hM$. To finalize this description, one needs
to specify the physical sense of functions of these operators appearing in Eqs.~(\ref{H_Init_A1})-(\ref{H_Xt}).

As we shall see below, operators~$\hO$ and~$\hM$ have the same eigenfunctions, so they commute.
Then the product of two exponential functions in Eqs.~(\ref{H_Init_A1})-(\ref{H_Init_Ap2})
is given by the exponential function with the sum of two exponents.
In consequence, there appear {\it two sets of frequencies}~$\omega^+$ and~$\omega^-$
corresponding to the sum and the difference:~$\omega^- \sim \hM-\hO$,
and~$\omega^+ \sim \hM+\hO$, respectively. The first frequencies~$\omega^-$, being of the intraband type,
lead in the non-relativistic limit to the cyclotron frequency~$\omega_c$.
The interband frequencies~$\omega^+$ correspond to the Zitterbewegung.
The electron motion is a sum of different frequency components when it is averaged over a wave packet.
In absence of a magnetic field {\it there are no intraband frequencies and only one interband frequency}
of the order of $2mc^2/\hbar$, see~\cite{Schroedinger1930}.

Each of the operators~$\hAD(t)$ or~$\hApD(t)$ contains both intraband and interband terms.
One could infer from Eqs.~(\ref{H_Init_A2}) and~(\ref{H_Init_Ap2}) that the amplitudes of
interband and intraband terms are similar. However,
when the explicit forms of the matrix elements of~$\hAD(t)$ and
$\hApD(t)$ are calculated, it will be seen that the ZB terms are much
smaller than the cyclotron terms, except at very high magnetic fields.

The operators~$\hO$ and~$\hM$ do not commute with~$\hAD$ or~$\hApD$. In
Eq.~(\ref{H_Init_A2}) the operator~$\hAD$ acts on the exponential terms from the right-hand side,
while in Eq.~(\ref{H_Init_Ap2}) the operator~$\hApD$ acts from the left-hand side.
The proper order of operators is to be retained in further calculations involving~$\hAD(t)$ or~$\hApD(t)$.

Let us consider the operator $\hM^2 = \hO^2 + \omega^2$.
Let $E_{\rm n}/\hbar$ and $|{\rm n}\rangle$ be the eigenvalue and eigenvector of~$\hO$,
respectively. Then
\begin{eqnarray} \label{M_M2}
\hM^2 |{\rm n}\rangle &=& (\hO^2+\omega^2)|{\rm n}\rangle
   = \frac{1}{\hbar^2}\left(E_{\rm n}^2 + \hbar^2\omega^2 \right) |{\rm n}\rangle.
\end{eqnarray}
Thus, every state $|{\rm n}\rangle$ is also an eigenstate of the operator~$\hM^2$
with the eigenvalue $\lambda_{\rm n}^2=E_{\rm n}^2/\hbar^2+\omega^2$. To find a more convenient
form of~$\lambda_{\rm n}$ we must find an explicit form of~$E_{\rm n}$. To do this we choose again
the Landau gauge ${\bm A} =(-By,0,0)$. Then, the eigenstate $|{\rm n}\rangle$ is
characterized by five quantum numbers: $n,k_x,k_z,\epsilon,s$,
where~$n$ is the harmonic oscillator number,~$k_x$ and~$k_z$ are the wave vectors in~$x$ and~$z$
directions, respectively,~$\epsilon=\pm 1$ labels the positive and negative energy branches,
and~$s=\pm 1$ is the spin index.
In the representation of Johnson and Lippman~\cite{Johnson1949} the state $|{\rm n}\rangle$ is
\begin{equation} \label{M_Lippman}
|{\rm n}\rangle =  N_{n\epsilon p_z}\left(\begin{array}{rl}
   s_1(\epsilon E_{n,k_z}+mc^2)&|n-1\rangle \\
   s_2(\epsilon E_{n,k_z}+mc^2)&|n\rangle \\
  (s_1p_zc -s_2\hbar \omega_n)&|n-1\rangle   \\
  -(s_1\hbar \omega_n + s_2 p_zc)&|n\rangle \end{array}\right),
\end{equation}
where~$s_1=(s+1)/2$ and~$s_2=(s-1)/2$ select the states~$s=\pm 1$, respectively.
The frequency is~$\omega_n=\omega \sqrt{n}$, the energy is
\begin{equation} \label{M_Enkz}
 E_{n,k_z}=\sqrt{(mc^2)^2 + (\hbar\omega_n)^2 + (\hbar k_zc)^2 },
\end{equation}
and the norm is $N_{n\epsilon k_z}=(2E_{n,k_z}^2+2\epsilon mc^2E_{n,k_z})^{-1/2}$.
In this representation the energy~$E_{n,k_z}$ does not depend explicitly on~$s$. Then the
eigenvalue of operator~$\hO$ is~$E_{\rm n}=\epsilon E_{n,k_z}/\hbar$.
The harmonic oscillator states are
\begin{equation} \label{M_nkxkz}
 \langle {\bm r} |n\rangle = \frac{e^{ik_xx+ik_zz}}{2\pi\sqrt{L}C_n}{\rm H}_{n}(\xi)e^{-1/2\xi^2},
\end{equation}
where~${\rm H}_{n}(\xi)$ are the Hermite polynomials and $C_n=\sqrt{2^n n!\sqrt{\pi}}$.
Using the above forms for $|{\rm n}\rangle$ and~$E_{n,k_z}$ we obtain from Eq.~(\ref{M_M2})
\begin{equation}
\hM^2 |{\rm n}\rangle =\frac{1}{\hbar^2} E_{n+1,k_z}^2 |{\rm n}\rangle,
\end{equation}
i.e. $\lambda_{\rm n}=\lambda_{n,k_z} = \pm E_{n+1,k_z}/\hbar$. In further calculations we assume
$\lambda_{n,k_z}$ to be positive. The operator~$\hM^2$ is diagonal. As follows from Eq.~(\ref{M_M2})
the explicit form of~$\hM^2$ is
\begin{equation}
 \hM^2 = {\rm diag} [d_1,d_2,d_1,d_2]
\end{equation}
with
\begin{eqnarray}
\hbar^2 d_1 &=&(mc^2)^2 + (cp_z^2) + \hbar^2\omega^2 +\hbar^2\omega^2\ha\hap,  \\
\hbar^2 d_2 &=&(mc^2)^2 + (cp_z^2) + \hbar^2\omega^2 +\hbar^2\omega^2\hap\ha.
\end{eqnarray}
Because~$\hM^2=\Omega^2+\omega^2$, eigenstates of~$\hM^2$ do
not depend on the energy branch index~$\epsilon$.

To calculate functions of operators~$\hO$ and~$\hM$ we use the fact that, for every reasonable
function~$f$ of operators~$\hO$ or~$\hM^2$, there is
$f(\hO)=\sum_{\rm n} f(\epsilon E_{n,k_z}/\hbar)|{\rm n}\rangle \langle {\rm n}|$,
and $f(\hM^2)=\sum_{\rm n} f(E_{n+1,k_z}^2/\hbar^2)|{\rm n}\rangle \langle {\rm n}|$,
see e.g~\cite{FeynmannBook}. Thus
\begin{eqnarray} \label{M_fM}
e^{\pm i\hO t}&=& \sum_{\rm n} e^{\pm i\epsilon t E_{n,k_z}/\hbar}|{\rm n}\rangle \langle {\rm n}|,        \\
\hM   &=& (\hM^2)^{1/2} = \nu\sum_{\rm n} \frac{ E_{n+1,k_z}}{\hbar }|{\rm n}\rangle \langle {\rm n}|,     \\
\hM^{-1} &=& (\hM^2)^{-1/2} = \nu\sum_{\rm n} \frac{\hbar}{E_{n+1,k_z}}|{\rm n}\rangle \langle {\rm n}|,   \\
e^{\pm i\hM t}&=& e^{\pm it(\hM^2)^{1/2}}\! = \!
          \sum_{\rm n}  e^{\pm i\nu tE_{n+1,k_z}/\hbar}|{\rm n}\rangle\langle {\rm n}|,
\end{eqnarray}
where~$\nu=\pm 1$. Without loss of generality we take~$\nu=+1$.
The above formulas can be used in calculating the matrix elements of~$\hAD(t)$ and~$\hApD(t)$.

Taking the eigenvectors~$|{\rm n}\rangle=|n,k_x,k_z,\epsilon,s\rangle$
and $|{\rm n'}\rangle=|n',k_x',k_z',\epsilon',s' \rangle$ with~$n'=n+1$,
we calculate matrix elements~$\hAD_{\rm n,n'}(t)$
using~$\hAD(t)$ given in Eqs.~(\ref{H_Init_A1}) and~(\ref{H_Init_A2}).
The selection rules for $\hAD_{\rm n,n'}(0)$ are~$k_x=k_x'$, and~$k_z=k_z'$, while $\epsilon,\epsilon',s,s'$
do not obey any selection rules. The matrix element of $\hM^{-1}\hAD(0)\hO$ appearing in
Eqs.~(\ref{H_Init_A1}) and~(\ref{H_Init_A2}) is
\begin{equation}
\langle {\rm n}|\hM^{-1}\hAD(0)\hO|{\rm n'}\rangle= \frac{1}{\lambda_{n,k_z}}\hAD(0)_{\rm n,n'}
 \frac{\epsilon' E_{n',k_z}}{\hbar} = \epsilon' \hAD(0)_{\rm n,n'}.
\end{equation}
In the last equality we used $E_{n',k_z}=E_{n+1,k_z}$ and $\lambda_{n,k_z}=E_{n+1,k_z}/\hbar$.
Introducing $\omega_{n,k_z}=E_{n,k_z}/\hbar$ we obtain
\begin{eqnarray}\label{M_AD1nnp}
\hAD_1(t)_{\rm n,n'} &=&\frac{1}{2}e^{i(\epsilon \omega_{n,k_z}-\lambda_{n,k_z})t}(1+\epsilon')\hAD(0)_{\rm n,n'}, \\
                 \label{M_AD2nnp}
\hAD_2(t)_{\rm n,n'} &=&\frac{1}{2}e^{i(\epsilon \omega_{n,k_z}+\lambda_{n,k_z})t}(1-\epsilon')\hAD(0)_{\rm n,n'}.
\end{eqnarray}
Thus the matrix element of $\hAD(t)_{\rm n,n'} = \hAD_1(t)_{\rm n,n'} + \hAD_2(t)_{\rm n,n'}$
is the sum of two terms, of which the first is nonzero
for~$\epsilon'=+1$, while the second is nonzero for~$\epsilon'=-1$.
As shown in Appendix~\ref{AppendixCheck}, the
matrix elements obtained in Eqs.~(\ref{M_AD1nnp}) and~(\ref{M_AD2nnp}) are equal to the
matrix elements of the Heisengerg operator
$\hAD(t)_{\rm n,n'}=\langle {\rm n}|e^{i\Omega t}\hAD(0)e^{-i\Omega t}|{\rm n'}\rangle$.

For $\hApD(t)_{\rm n',n}=\hApD_1(t)_{\rm n',n}+\hApD_2(t)_{\rm n',n}$ we obtain in a similar way
\begin{eqnarray}\label{M_ApD1nnp}
\hApD_1(t)_{\rm n',n} &=&\frac{1}{2}e^{i(+\lambda_{n,k_z}-\epsilon \omega_{n,k_z})t}(1+\epsilon')\hApD(0)_{\rm n',n}, \ \ \\
                \label{M_ApD2nnp}
\hApD_2(t)_{\rm n',n} &=&\frac{1}{2}e^{i(-\lambda_{n,k_z}-\epsilon \omega_{n,k_z})t}(1-\epsilon')\hApD(0)_{\rm n',n}. \ \
\end{eqnarray}
Formulas~(\ref{M_AD1nnp})-(\ref{M_ApD2nnp}) describe the time evolution of the matrix elements of~$\hAD(t)$
and~$\hApD(t)$ calculated between two eigenstates of~$\hO$. The frequencies appearing in the exponents are of the
form $\pm \lambda_{n,k_z} \pm \omega_{n,k_z} = \pm \omega_{n+1,k_z} \pm \omega_{n,k_z}$. The intraband terms
characterized by $\omega_n^c =\omega_{n+1,k_z} - \omega_{n,k_z}$
correspond to the cyclotron motion, while the interband terms characterized by
$\omega_n^Z =\omega_{n+1,k_z} + \omega_{n,k_z}$
describe ZB. Different values of $\epsilon, \epsilon'$ in the
matrix elements of $\hAD_1(t)_{\rm n,n'}, \hAD_2(t)_{\rm n',n},
\hApD_1(t)_{\rm n',n}, \hApD_2(t)_{\rm n',n}$ give contributions either to the cyclotron or to the ZB motion.
In Appendix~\ref{AppendixCheck} we tabulate the above matrix elements for all combinations of $\epsilon, \epsilon'$.
The exact compact results given in Eqs.~(\ref{M_AD1nnp})-(\ref{M_ApD2nnp}) indicate that our
choice of~$\hAD(t)$ and~$\hApD(t)$ operators for the description of relativistic electrons
in a magnetic field was appropriate.

To complete the operator considerations of ZB we estimate
low-field and high-field limits of $\hAD_{\rm n,n'}(t)$.
Consider first the matrix element between two states of positive energies and~$s=-1$. We take
$|{\rm n}\rangle=|n,k_x,k_z,+1,-1\rangle$ and $|{\rm n'}\rangle=|n+1,k_x,k_z,+1,-1\rangle$. Then
\begin{eqnarray} \label{M_Limit_1}
\hAD(t)_{\rm n,n'}^{c}=\sqrt{n+1}\ \ e^{i(E_{n,k_z}-E_{n+1,k_z})t/\hbar} \times \nonumber \\
   \frac{(E_{n,k_z}+E_{n+1,k_z})(E_{n,k_z}+mc^2)}{2\sqrt{E_{n,k_z}E_{n+1,k_z}(E_{n,k_z}+mc^2)(E_{n+1,k_z}+mc^2)}}.
\end{eqnarray}
This equals $\hAD_1(t)_{\rm n,n'}$ given in Eq.~(\ref{M_AD1nnp}) because $\hAD_2(t)_{\rm n,n'}=0$ for~$\epsilon'=+1$.
At low magnetic fields there is
\begin{equation}
E_{n+1,k_z} - E_{n,k_z} = \frac{\hbar^2\omega^2}{E_{n+1,k_z} + E_{n,k_z}} \simeq \frac{\hbar eB}{m}\equiv \hbar\omega_c,
\end{equation}
where in the denominators we approximated $E_{n,k_z}\simeq E_{n+1,k_z} \simeq mc^2$
and used $\omega=\sqrt{2}c/L$, see Eq.~(\ref{H aap}). Setting again $E_{n,k_z}\simeq E_{n+1,k_z} \simeq mc^2$ in
the numerator and denominator of Eq.~(\ref{M_Limit_1}) we recover the well known result for the
matrix elements of the lowering operator~$\ha$ in the non-relativistic limit
\begin{equation}
\hAD(t)_{\rm n,n'}\simeq \sqrt{n+1}\ e^{-i\omega_ct}.
\end{equation}
Consider now the above state $|{\rm n}\rangle$ from the positive energy branch
and the state $|{\rm n'}\rangle$ from the negative energy branch
$|{\rm n'}\rangle =|n+1,k_z,k_z,-1,-1\rangle$. Then the matrix element is
\begin{eqnarray}
\hAD(t)_{\rm n,n'}=\sqrt{n+1}\ \ e^{i(E_{n,k_z}+E_{n+1,k_z})t/\hbar} \times \nonumber \\
   \frac{(E_{n,k_z}-E_{n+1,k_z})(E_{n,k_z}+mc^2)}{2\sqrt{E_{n,k_z}E_{n+1,k_z}(E_{n,k_z}+mc^2)(E_{n+1,k_z}-mc^2)}}.
\end{eqnarray}
Assuming low magnetic fields, small~$k_{0z}$ values, and using the above approximations we obtain
\begin{equation}
\hAD(t)_{\rm n,n'}^{ZB}\simeq \sqrt{\frac{\hbar \omega_c}{2mc^2}} e^{-2imc^2t/\hbar}.
\end{equation}
Since at low magnetic fields there is $\hbar \omega_c \ll mc^2$, the amplitude of interband~(Zitterbewegung)
oscillations is much lower than that of the cyclotron motion. At low magnetic fields both the amplitude
and the frequency of ZB do not depend on the quantum number~$n$.

Let us consider now the opposite case of very strong magnetic fields, when $\hbar \omega \gg mc^2$
and $\hbar \omega \gg \hbar ck_z$. Such a situation is difficult to realize experimentally
since the condition $\hbar \omega = mc^2$ corresponds to $L=\sqrt{2}\lambda_c$, i.e. the magnetic
length is of the order of the Compton wavelength. Within this limit $E_{n,k_z}\simeq E_n=\hbar\omega\sqrt{n}$,
and the matrix elements of $\hAD(t)_{\rm n,n'}$ for the cyclotron and ZB components are
\begin{equation}
\hAD(t)_{\rm n,n'}=(\sqrt{n}\ \pm \sqrt{n+1}\ )e^{i\omega(\sqrt{n}\ \mp \sqrt{n+1}\ )t},
\end{equation}
where the upper signs corresponds to the cyclotron and the lower ones to the ZB motion, respectively.

The conclusion from the above analysis is that at low magnetic fields of a few tenths
of Tesla the ZB amplitude is eight orders of magnitude smaller than the cyclotron amplitude. In fields of the
order of $4.4\times 10^{9}$ T the ZB motion and cyclotron motion are of the same orders of magnitude.
This completes our derivation and analysis of the operators describing electron
motion in a magnetic field according to the 'empty' Dirac equation.
However, it is well known that observable quantities are given by average values.

\section{Zitterbewegung: average values}
In this section we concentrate on observable quantities, i.e. on electron positions and velocities
averaged over a wave packet~$f({\bm r})$. We analyze the one-electron Dirac equation
neglecting many-body effects.
Our calculations are first performed for a general form of~$f({\bm r})$ and then specialized for the Gaussian
form of the packet.

\subsection{Averaging procedure}
We take a packet with one or two nonzero components, i.e. $f({\bm r})(a_1,a_2,0,0)^T$
with $|a_1|^2+|a_2|^2=1$.
According to the procedure adopted in the previous section, we first calculate the
averages of~$\hAD(t)$ and~$\hApD(t)$ operators and then the position operators~$\hY(t)$
and~$\hX(t)$.
We do not consider multi-component packets because they are difficult to prepare and their
physical sense is not clear.

Averaging of operators~$\hAD(t)$ and~$\hApD(t)$ can be performed using formulas
from the previous section, see Eqs.~(\ref{H_Init_A1})-(\ref{H_Init_Ap2}).
However, a simpler and more general method is to average the Heisenberg time-dependent form
 $\hAD(t)=e^{i\hO t}\hAD(0) e^{-i\hO t}$
with the use of two unity operators $\hat{\bm 1} = \sum_{\rm n}|{\rm n}\rangle\langle {\rm n}|$.
Then the average of~$\hAD(t)$ is
\begin{eqnarray} \label{Avg_A}
\langle \hAD(t)\rangle = \langle f|\hAD(t)|f\rangle =
\langle f|e^{i\hH t/\hbar}\hAD e^{-i\hH t/\hbar}|f\rangle = \nonumber \\
 \sum_{\rm nn'} e^{i\epsilon E_{\rm n,k_z}t/\hbar} e^{-i\epsilon' E_{\rm n'k_z}t/\hbar }
           \langle {\rm n}|f\rangle  \langle f| {\rm n'}\rangle \langle {\rm n}|\hAD|  {\rm n'} \rangle,
\end{eqnarray}
and similarly for~$\langle\hApD(t)\rangle$. There is
\begin{equation}
\sum_{\rm nn'} \Rightarrow \sum_{n,n'}\sum_{\epsilon \epsilon'} \sum_{s,s'} \int dk_x dk_x' dk_z dk_z'.
\end{equation}
The selection rules for the matrix elements $\langle {\rm n}|\hAD|  {\rm n'} \rangle$ are:
$n'=n+1$, $k_x'=k_x$, $k_z'=k_z$, while for $\langle {\rm n}|\hApD| {\rm n'} \rangle$
we have $n'=n-1$, $k_x'=k_x$, $k_z'=k_z$. The wave packet is assumed to be separable
$f({\bm r}) = f_z(z)f_{xy}(x,y)$. Then we have
\begin{equation} \label{Avg_Pak_nf}
\langle {\rm n}|f\rangle = \chi_{n\epsilon k_z} g_z(k_z) (s_1 a_1F_{n-1} + s_2 a_2F_{n}),
\end{equation}
where $\chi_{n\epsilon k_z}=(\epsilon E_{n,k_z} + mc^2) N_{n\epsilon k_z}$, and
\begin{equation} \label{Avg_Fn}
 F_n(k_x) = \frac{1}{\sqrt{2L}C_n} \int_{-\infty}^{\infty} g_{xy}(k_x,y)e^{-\frac{1}{2}\xi^2}{\rm H}_{n}(\xi)dy,
\end{equation}
in which
\begin{equation}
 g_{xy}(k_x,y) = \frac{1}{\sqrt{2\pi}} \int_{-\infty}^{\infty} f_{xy}(x,y)e^{ik_xx} dx,
\end{equation}
and
\begin{equation}
 g_z(k_z) = \frac{1}{\sqrt{2\pi}} \int_{-\infty}^{\infty} f_{z}(z)e^{ik_zz} dz.
\end{equation}
To proceed further we must specify nonzero components $a_1, a_2$ of the wave packet.
First, we limit our calculations to a one-component packet with the nonzero
second component corresponding to the state with the spin~$s_z=-1/2$.
Setting $a_1=0, a_2=1$ we obtain from Eq.~(\ref{Avg_Pak_nf}):
$\langle {\rm n}|f\rangle = s_2 \chi_{n\epsilon k_z} g_z(k_z) F_{n}(k_z)$.
This gives
\begin{eqnarray}
\langle \hAD(t)\rangle^{2,2} =
\sum_{n,n'}\int_{-\infty}^{\infty} dk_z dk_z' g_z^*(k_z) g_z(k_z') \times \nonumber \\
\sum_{\epsilon,\epsilon'}
e^{i(\epsilon E_{n,k_z}-\epsilon' E_{n'k_z})t/\hbar} \chi_{n\epsilon k_z}\chi_{n^{'}\epsilon' k_z'}
\times \nonumber \\
 \int_{-\infty}^{\infty} dk_x dk_x' F_{n}^*(k_x) F_{n'}(k_x')
 \sum_{s,s'} s_2 s_2' \langle {\rm n}|\hAD| {\rm n'} \rangle.
\end{eqnarray}
The upper indices in $\langle \hAD(t)\rangle^{2,2}$ indicate the second nonzero component of
the wave packet involved.
The matrix element $\langle {\rm n}|\hAD| {\rm n'} \rangle^{2,2}$
has ten nonzero terms. The summation $\sum_{s's} s_2s_2' \langle {\rm n}|\hAD| {\rm n'} \rangle$
gives only three nonzero terms being the products of $(s_2s_2')^2$, since
$s_1s_2=s_1's_2'=0$. Rearranging summations and integrations we obtain
\begin{eqnarray}
\lefteqn{\langle \hAD(t)\rangle^{2,2} =\sum_n U_{n,n+1}\sqrt{n+1}\ \times } \nonumber \\
&&\int_{-\infty}^{\infty} dk_z |g_z(k_z)|^2 \sum_{\epsilon,\epsilon'}
e^{i(\epsilon E_{n,k_z}-\epsilon' E_{n+1,k_z})t/\hbar}\times
\nonumber \\
&& \left[\chi_{n\epsilon k_z}^2\chi_{n+1\epsilon k_z}^2 +
  \eta_{n\epsilon k_z}\eta_{n+1,\epsilon k_z}(c^2p_z^2+\hbar^2\omega_{n}^2) \right], \ \ \ \
\end{eqnarray}
where $\eta_{n\epsilon k_z} = \chi_{n\epsilon k_z} N_{n\epsilon k_z}$.
We define
\begin{equation} \label{Avg_Umn}
 U_{m,n} = \int_{-\infty}^{\infty} F_{m}^*(k_x) F_{n}(k_x) dk_x.
 \end{equation}
Since $\chi_{n\epsilon k_z}^2=(1/2)+\epsilon mc^2/(2E_{n,k_z})$,
      $\eta_{n\epsilon k_z}=\epsilon/(2E_{n,k_z})$, and $E_{n,k_z}^2=(mc^2)^2+(cp_z)^2+ (\hbar\omega_{n})^2$,
we have
\begin{eqnarray} \label{Avg_22_A_0}
\lefteqn{\langle \hAD(t)\rangle^{2,2} =\sum_n U_{n,n+1}\sqrt{n+1}\ \times} \nonumber \\
&&\int_{-\infty}^{\infty} dk_z |g_z(k_z)|^2 \sum_{\epsilon,\epsilon'}
e^{i(\epsilon E_{n,k_z}-\epsilon' E_{n+1,k_z})t/\hbar}\times
\nonumber \\
&& \frac{1}{4}\left[1 + \epsilon\epsilon'\frac{E_{n,k_z}}{E_{n+1,k_z}}
        + mc^2\left(\frac{\epsilon}{E_{n,k_z}} + \frac{\epsilon'}{E_{n+1,k_z}}\right) \right]. \ \ \
\end{eqnarray}
The summations over~$\epsilon$ and~$\epsilon'$ lead to combinations of sine and cosine functions.
The calculation of $\langle\hApD(t)\rangle$ is similar to that shown above, but the selection
rules for $\langle {\rm n}|\hApD| {\rm n'} \rangle$ are $n'=n-1$, $k_x'=k_x$, $k_z'=k_z$.
Performing the summations we finally obtain
\begin{eqnarray}
 \label{Avg_22_At} \langle\hAD(t) \rangle^{2,2} \!\! &=&\!\! \frac{1}{2}\sum_n \sqrt{n+1}\ U_{n,n+1}
      \times \nonumber \\ && \left(I^+_c + I^-_c - iI^+_s - iI^-_s \right), \\
 \label{Avg_22_Apt} \langle\hApD(t)\rangle^{2,2} \!\! &=&\!\! \frac{1}{2}\sum_n \sqrt{n+1}\ U_{n+1,n}
      \times \nonumber \\ && \left(I^+_c + I^-_c + iI^+_s + iI^-_s \right),
\end{eqnarray}
where
\begin{eqnarray}
    \label{Avg_Ic}
 I^{\pm}_c &=& \int_{-\infty}^{\infty}\left(1 \pm \frac{E_{n,k_z}}{E_{n+1,k_z}} \right)|g_z(k_z)|^2 \times \nonumber \\
           && \ \ \ \ \ \ \cos\left[(E_{n+1,k_z} \mp E_{n,k_z})t/\hbar \right]dk_z, \\
    \label{Avg_Is}
 I^{\pm}_s &=& mc^2\int_{-\infty}^{\infty} \left(\frac{1}{E_{n,k_z}} \pm \frac{1}{E_{n+1,k_z}} \right)
           |g_z(k_z)|^2 \times \nonumber \\
           && \ \ \ \ \ \ \sin\left[(E_{n+1,k_z} \mp E_{n,k_z})t/\hbar \right]dk_z.
\end{eqnarray}
Finally, average electron positions $\langle \hY(t)\rangle^{2,2}$ and $\langle \hX(t)\rangle^{2,2}$
for the 3+1 Dirac equation in a vacuum are
[see Eqs.~(\ref{Avg_22_At}) and~(\ref{Avg_22_Apt}), and Eqs.~(\ref{H_Yt}) and~(\ref{H_Xt})]
\begin{eqnarray}
\label{Avg_22_Yt} \langle \hY(t)\rangle^{2,2} &=& \frac{L}{2\sqrt{2}} \sum_n \sqrt{n+1}\ \times \nonumber \\
  && \left(U_{n,n+1}+U_{n+1,n}\right) \left(I^+_c + I^-_c \right) + k_{0x}L^2, \ \ \ \ \ \\
\label{Avg_22_Xt} \langle \hX(t)\rangle^{2,2} &=& \frac{L}{2\sqrt{2}} \sum_n \sqrt{n+1}\ \times \nonumber \\
  && \left(U_{n,n+1}+U_{n+1,n}\right) \left(I^+_s + I^-_s \right).
\end{eqnarray}

For a packet with the first nonzero component we obtain similar results.
In both cases there appear {\it the same frequencies} but they enter to the motion with different amplitudes.
This is illustrated in Fig. \ref{FigSim3} of Section V for the 2+1 Dirac equation.
The averages $\langle \hY(t)\rangle$ and $\langle \hX(t)\rangle$ are equal, up to a constant
$y_0=k_{0x}L^2$, to the averages of the usual position
operators $\langle \hat{y}(t)\rangle$ and $\langle \hat{x}(t)\rangle$, see Appendix~\ref{AppendixXY}.

Finally we consider a two-component wave packet
$\langle {\bm r} | f\rangle = f({\bm r}) (a_1,a_2, 0,0)^T$ with $|a_1|^2+|a_2|^2=1$.
Defining~$f_1=a_1 f$ and~$f_2=a_2 f$ we have
\begin{eqnarray}
\langle f_1 + f_2|\hAD(t)| f_1+ f_2\rangle =
\langle f_1 |\hAD(t)| f_1\rangle + \langle f_2|\hAD(t)| f_2\rangle + \nonumber \\
\langle f_1 |\hAD(t)| f_2\rangle + \langle f_2|\hAD(t)| f_1\rangle, \ \ \ \ \
\end{eqnarray}
and similarly for $\langle f_1 + f_2|\hApD(t)| f_1+ f_2\rangle$.
The first two terms were calculated above. The other two terms are
\begin{eqnarray}
\langle\hAD(t) \rangle^{2,1}= \frac{1}{2}a_2^*a_1\sum_n U_{n,n} \left(J^+_c    + J^-_c    \right), \\
\langle\hApD(t)\rangle^{1,2}= \frac{1}{2}a_1^*a_2\sum_n U_{n,n} \left(J^{+*}_c + J^{-*}_c \right),
\end{eqnarray}
and $\langle\hAD(t)\rangle^{1,2}= \langle\hApD(t)\rangle^{2,1} = 0$. We define
\begin{eqnarray} \label{Avg_J}
J^{\pm}_c &=& \pm \int_{-\infty}^{\infty} \frac{cp_z\hbar\omega}{E_{n,k_z}E_{n+1,k_z}}
  g_z^*(k_z)g_z(k_z) \times \nonumber \\
           && \ \ \ \ \ \ \cos\left[(E_{n+1,k_z} \mp E_{n,k_z})t/\hbar \right]dk_z.
\end{eqnarray}
The integrals~$J^{\pm}_c$ describe mixing of the
states with different components~$s_z$. Since~$J^{\pm}_c$ are odd functions
of~$k_z$, they vanish for the wave packet with~$k_{0z}=0$. Contributions from these integrals are
relevant only for magnetic fields of the order of $B\simeq 5 \times 10^{9}$ T,
where the magnetic length~$L$ is comparable to~$\lambda_c$.
The velocity of the packet in the~$z$ direction $v_z=\hbar k_{0z}/m$ must be comparable to~$c$.
At low magnetic fields the mixing terms are negligible.

All the above results were obtained the for the 3+1 Dirac equation. A reduction to the 2+1 DE
is obtained by setting $|g(k_z)|^2=\delta(k_z)$ in Eqs.~(\ref{Avg_Ic}) and~(\ref{Avg_Is})
and performing integrations over~$k_z$. Below we quote
final results for $\langle \hY(t)\rangle^{2,2}$
and $\langle \hX(t)\rangle^{2,2}$ for the latter case
\begin{widetext}
\begin{eqnarray}        \label{Avg_22_Yt_2D}
\langle \hY(t)\rangle^{2,2} &=& \frac{L}{2\sqrt{2}} \sum_n \sqrt{n+1}\
   \left(U_{n,n+1}+U_{n+1,n}\right)
       \left\{ \left(1+\frac{E_{n}}{E_{n+1}} \right) \cos(\omega_n^ct) +
               \left(1-\frac{E_{n}}{E_{n+1}} \right) \cos(\omega_n^Zt)\right\} + k_{0x}L^2, \\
\label{Avg_22_Xt_2D}
\langle \hX(t)\rangle^{2,2} &=& \frac{L}{2\sqrt{2}} \sum_n \sqrt{n+1}\
\left(U_{n,n+1}+U_{n+1,n}\right)
       \left\{ \left(\frac{mc^2}{E_{n}}+\frac{mc^2}{E_{n+1}} \right) \sin(\omega_n^ct) +
               \left(\frac{mc^2}{E_{n}}-\frac{mc^2}{E_{n+1}} \right) \sin(\omega_n^Zt)\right\}.
\end{eqnarray}
\end{widetext}
In the above equations we used notation $E_n\equiv E_{n,k_z=0}$, $\omega_n^c=(E_{n+1}-E_{n})/\hbar$
and $\omega_n^Z=(E_{n+1}+E_{n})/\hbar$.
For the 2+1 Dirac equation the final expressions for $\langle \hY(t)\rangle^{2,2}$
and $\langle \hX(t)\rangle^{2,2}$ are given in form of infinite sums,
while for the 3+1 DE they are given by infinite sums and integrals over~$k_z$.
As is known from the Riemann-Lesbegues theorem (see Ref.~\cite{Lock1979}),
the $k_z$ integrals over rapidly oscillating functions of time,
appearing in Eqs.~(\ref{Avg_Ic}) and~(\ref{Avg_Is}),
decay to zero after sufficiently long times. Therefore, the packet motion for the 3+1 Dirac equation
has a transient character, while that for the 2+1 DE is persistent. Transient and persistent ZB motions
in the two cases are illustrated in Fig.~\ref{FigSim4} of Section V.

\subsection{Gaussian wave packet}
We perform specific calculations for one- or two-component wave packets taking
the function~$f({\bm r})$ in form of an ellipsoidal Gaussian packet
characterized by three widths $d_x$, $d_y$, $d_z$ and having a nonzero momentum
$\hbar {\bm k}_0=\hbar(k_{0x},0,k_{0z})$
\begin{equation} \label{Gauss_packet}
f({\bm r}) = \frac{1}{\sqrt{\pi^3d_x d_y d_z}}\exp
   \left(-\frac{x^2}{2d_x^2}-\frac{y^2}{2d_y^2}-\frac{z^2}{2d_z^2}
      + i{\bm k}_0 {\bm r} \right).
\end{equation}
The wave packet is multiplied by a four-component Dirac spinor $(a_1,a_2,0,0)^T$.
Using the definitions of $g_{xy}(k_x,y)$, $F_n(k_x)$ and $U_{m,n}$,
we obtain (see Refs.~\cite{Rusin20008a,PrudnikovBook})
\begin{equation}
g_{xy}(k_x,y) = \sqrt{\frac{d_x}{\pi d_y}}e^{-\frac{1}{2}d_x^2(k_x-k_{0x})^2} e^{-\frac{y^2}{2d_y^2}}
\end{equation}
and
\begin{equation} \label{Gauss_Fn}
 F_n(k_x) = \frac{A_n\sqrt{L d_x}}{\sqrt{2\pi d_y}C_n} e^{-\frac{1}{2}d_x^2(k_x-k_{0x})^2}
            e^{-\frac{1}{2}k_x^2D^2}\ {\rm H}_n(-k_xc),
\end{equation}
where $D=L^2/\sqrt{L^2+d_y^2}$, $c=L^3/\sqrt{L^4-d_y^4}$, and
\begin{equation}
 A_n=\frac{\sqrt{2\pi}d_y}{\sqrt{L^2+d_y^2}}\left(\frac{L^2-d_y^2}{L^2+d_y^2}\right)^{n/2},
\end{equation}
\begin{eqnarray} \label{Gauss_Umn}
U_{m,n}= \frac{A_m^* A_nLQd_x \sqrt{\pi}\ e^{-W^2}}{\pi C_mC_n d_y}
    \sum_{l=0}^{\min\{m,n\}}\!\!\!\! 2^l l! \! \left(\begin{array}{c} m \\ l\!\! \end{array}\right) \!\!\!
                                         \left(\begin{array}{c} n \\ l\!\! \end{array}\right)
   &&    \nonumber \\
  \times (\left(1-(cQ)^2\right)^{(m+n-2l)/2}
    {\rm H}_{m+n-2l} \left(\frac{-cQY}{\sqrt{1-(cQ)^2}}\right), \ \ \ \ \ &&
\end{eqnarray}
in which $Q = 1/\sqrt{d_x^2+D^2}$, $W= d_xDQk_{0x}$, and $Y=d_x^2k_{0x}Q$.
For the special case of~$d_y=L$, the formula for~$U_{m,n}$ is much simpler:
\begin{eqnarray} \label{Gauss_Umn_Ldy}
U_{m,n} &=& 2\frac{\sqrt{\pi}\ (-i)^{m+n}\ d_x}{C_mC_nL}
   \left(\frac{L}{2P} \right)^{m+n+1} \times \nonumber \\
 && \exp\left(-\frac{d_x^2k_{0x}^2L^2}{2P^2}\right)
     {\rm H}_{m+n}\left(\frac{-id_x^2k_{0x}}{P}\right), \ \ \
\end{eqnarray}
where $P=\sqrt{d_x^2+\frac{1}{2}L^2}$. In the above expressions
the coefficients $U_{m,n}$ are real numbers and they are symmetric
in~$m,n$ indices. For further discussion of
of~$U_{m,n}$ see Appendix~\ref{AppendixUmn} and Ref.~\cite{Rusin20008a}.

The coefficients $U_{m,n}$ given in Eqs.~(\ref{Gauss_Umn}) and~(\ref{Gauss_Umn_Ldy}), apart
from the~$k_z$ dependent parts of the integrals~$I_c^{\pm}$ and~$I_s^{\pm}$, describe the amplitudes of
oscillation terms. In the special case of~$n=m$ they are the probabilities of the
expansion of a packet~$f({\bm r})$
in eigenstates of the Hamiltonian $\hat{H}=(\hbar^2/2m)(\hat{\bm p} -e{\bm A})^2$ of an electron
in a uniform magnetic field.
This ensures that all $U_{n,n}$ are non-negative and normalized to unity, so that
in practice there is a finite number of non-negligible~$U_{n,n}$ coefficients. There is also a summation rule
for $\sqrt{n+1}\ U_{n+1,n}$, see Appendix~\ref{AppendixUmn}, which reduces
the number of non-negligible coefficients $U_{n+1,n}$. Finite number of non-negligible
coefficients~$U_{n,m}$ limits the number of frequencies contributing to the cyclotron and ZB motions.
Simpler formula~(\ref{Gauss_Umn_Ldy}) for~$U_{m,n}$ shows that
the coefficients $U_{n,n+1}$ are relevant if all the quantities $d_x$, $d_y$, $k_{0x}^{-1}$
and the magnetic length~$L$ are
of the same order of magnitude. The remaining parameters, i.e. $d_z$ and $k_{0z}$,
can be arbitrary with the only requirement that the total initial packet
velocity $|{\bm v}_0|=\hbar|{\bm k}_0|/m$ must be
smaller than~$c$, which is equivalent to $\sqrt{k_{0x}^2+k_{0z}^2}<\lambda_c^{-1}$.
Because of the~$x-y$ symmetry of our problem, it is natural to take
$d_x \approx d_y$. In our calculations we keep
$d_x\approx d_y \approx d_z$, but they do not have to be equal. Because a constant magnetic
field does not create electron-hole pairs, there is no restriction on~$B$ and the magnetic length~$L$
can be arbitrarily small.

Before presenting numerical calculations for the motion of a wave packet in a magnetic field
we analyze qualitatively possible regimes of parameters for realistic physical situations.
This problem has two characteristic lengths: the Compton wavelength $\lambda_c=3.86 \times$10$^{-3}$\AA\
and the magnetic length~$L$. For a magnetic field $B=40$ T there is $L=40.6$\AA.
The magnetic length is equal to $\lambda_c$ for $B=4.4\times$ 10$^9$ T.
We then distinguish two regimes of parameters: i) the low-field limit, in which packet widths $d_x$, $d_y$,
$k_{0x}^{-1}$ and the magnetic length~$L$ are of the order of nanometers, and ii) the relativistic regime,
in which all quantities $d_x$, $d_y$, $k_{0x}^{-1}$ and~$L$ are of the order of~$\lambda_c$.

\subsection{Low magnetic fields}
At low magnetic fields the electron moves on a circular orbit with the frequency~$\omega_c=eB/m$ and
the radius~$r=mv/eB$. The aim of this subsection is to retrieve the non-relativistic cyclotron motion
from the general formulas in Eqs.~(\ref{Avg_22_Yt})-(\ref{Avg_22_Xt}). Additionally, we show
that ZB exists even at low magnetic fields but its amplitude is much smaller than~$\lambda_c$.

At low magnetic fields we approximate $E_{n,k_z}\simeq mc^2$ and $E_{n+1,k_z}-E_{n,k_z}\simeq \hbar\omega_c$.
Then~$I_c^-$ and~$I_s^-$ in Eqs.~(\ref{Avg_Ic}) and~(\ref{Avg_Is}) reduce to
\begin{eqnarray}
I_c^- = 2\cos(\omega_ct) \int_{-\infty}^{\infty} |g_z(k_z)|^2 dk_z, \\
I_s^- = 2\sin(\omega_ct) \int_{-\infty}^{\infty} |g_z(k_z)|^2 dk_z,
\end{eqnarray}
and they do not depend on~$n$. The integrals over~$k_z$ give unity due to the normalization of
the wave packet. The summation over~$n$ in Eqs.~(\ref{Avg_22_Yt})-(\ref{Avg_22_Xt}) is performed
with the use of the formula (see Appendix~\ref{AppendixUmn})
\begin{equation}
\sum_{n=1}^{\infty} \sqrt{n+1}\ U_{n+1,n} = -\frac{1}{\sqrt{2}}k_{0x}L.
\end{equation}
We find
\begin{eqnarray} \label{LFL_y}
 \langle y(t) \rangle &\simeq& -k_{0x}L^2\cos(\omega_ct) + k_{0x}L^2, \\
 \langle x(t) \rangle &\simeq& -k_{0x}L^2\sin(\omega_ct). \label{LFL_x}
\end{eqnarray}
Since $L^2=\hbar/eB$ and $v_{0x}=\hbar k_{0x}/m$, we obtain $k_{0x}L^2 = mv_{0x}/eB$,
which is equal to the radius of the cyclotron motion.
Taking the time derivative of $\langle y(t) \rangle$ and $\langle x(t) \rangle$
and using definitions of~$L$ and~$\omega_c$ we have
\begin{eqnarray}
\langle v_y(t)\rangle &\simeq& \frac{\hbar k_{0x}}{m}\sin(\omega_ct), \\
\langle v_x(t)\rangle &\simeq& -\frac{\hbar k_{0x}}{m}\cos(\omega_ct).
\end{eqnarray}
Thus we recover the cyclotron motion of a non-relativistic
electron in a constant magnetic field.

Now we turn to the ZB motion. At the low-field limit we again separate the integration over~$k_z$
from the summation over~$n$. The integration over~$k_z$ selects~$k_z\simeq 0$,
so the amplitude~$D$ of the ZB motion is
[see Eq.~(\ref{Avg_22_Yt})]
\begin{eqnarray} \label{LFL_D}
D &\simeq& \frac{L}{2\sqrt{2}} \left(1 - \frac{E_{n,0}}{E_{n+1,0}} \right) \sum_n \sqrt{n+1} \
  \left(U_{n+1,n} + U_{n,n+1}\right) \nonumber \\
  &\simeq& \frac{L}{2\sqrt{2}} \times \frac{\hbar^2\omega^2}{2m^2c^4}\times \frac{2k_{0x}L}{\sqrt{2}} =
 \frac{1}{2} \lambda_c (k_{0x}\lambda_c).
\end{eqnarray}
Thus at low magnetic fields the amplitude of the ZB motion is a small fraction of~$\lambda_c$, since
$k_{0x}\lambda_c \ll 1$. This agrees with the old predictions of Lock in Ref.~\cite{Lock1979}.
An interesting feature of ZB motion at low magnetic fields is its slow decay in time,
proportional to~$t^{-1/2}$. A similar decay of ZB proportional to $t^{-1/2}$ was also
predicted for a one-dimensional electron Zitterbewegung in carbon nanotubes~\cite{Rusin07b}.
To understand this behavior we consider the integral~$I^+_c(t)$
in Eq.~(\ref{Avg_Ic}). Retaining only the cosine function and taking a Gaussian wave packet we obtain
\begin{equation}
I^+_c(t) \simeq D^{'} \int_{-\infty}^{\infty}\cos\left[(E_{n+1,k_z}+E_{n,k_z})t/h\right] e^{-d_z^2k_z^2} dk_z,
\end{equation}
where~$D^{'}$ is a constant independent of~$k_z$ and proportional to~$D$, as given in Eq.~(\ref{LFL_D}).
Expanding the energy $E_{n,k_z}$ in Eq.~(\ref{M_Enkz}) to the lowest terms in~$k_z$, we have
\begin{equation}
I^+_c(t) \simeq D^{'}\int_{-\infty}^{\infty} \cos\left[\left(2 +k_z^2\lambda_c^2\right)
  \frac{mc^2t}{\hbar} \right] e^{-d_z^2k_z^2} dk_z.
\end{equation}
The direct integration gives
\begin{equation}
I^+_c(t) \simeq D^{'} \frac{F^{osc}(t)}{\left[d_z^4 + (\hbar t/m)^2 \right]^{1/4}},
\end{equation}
where~$F^{osc}(t)$ is a function oscillating with the frequency~$\omega=2mc^2/\hbar$ and having
the amplitude of the order of unity. Therefore, the ZB oscillations decay as~$t^{-1/2}$ and they
persist even at times of picoseconds. This is illustrated in Fig.~\ref{Fig3DLow} of Section IV.

\section{Results: 3+1 Dirac equation}
\begin{figure}
\includegraphics[width=8.5cm,height=8.5cm]{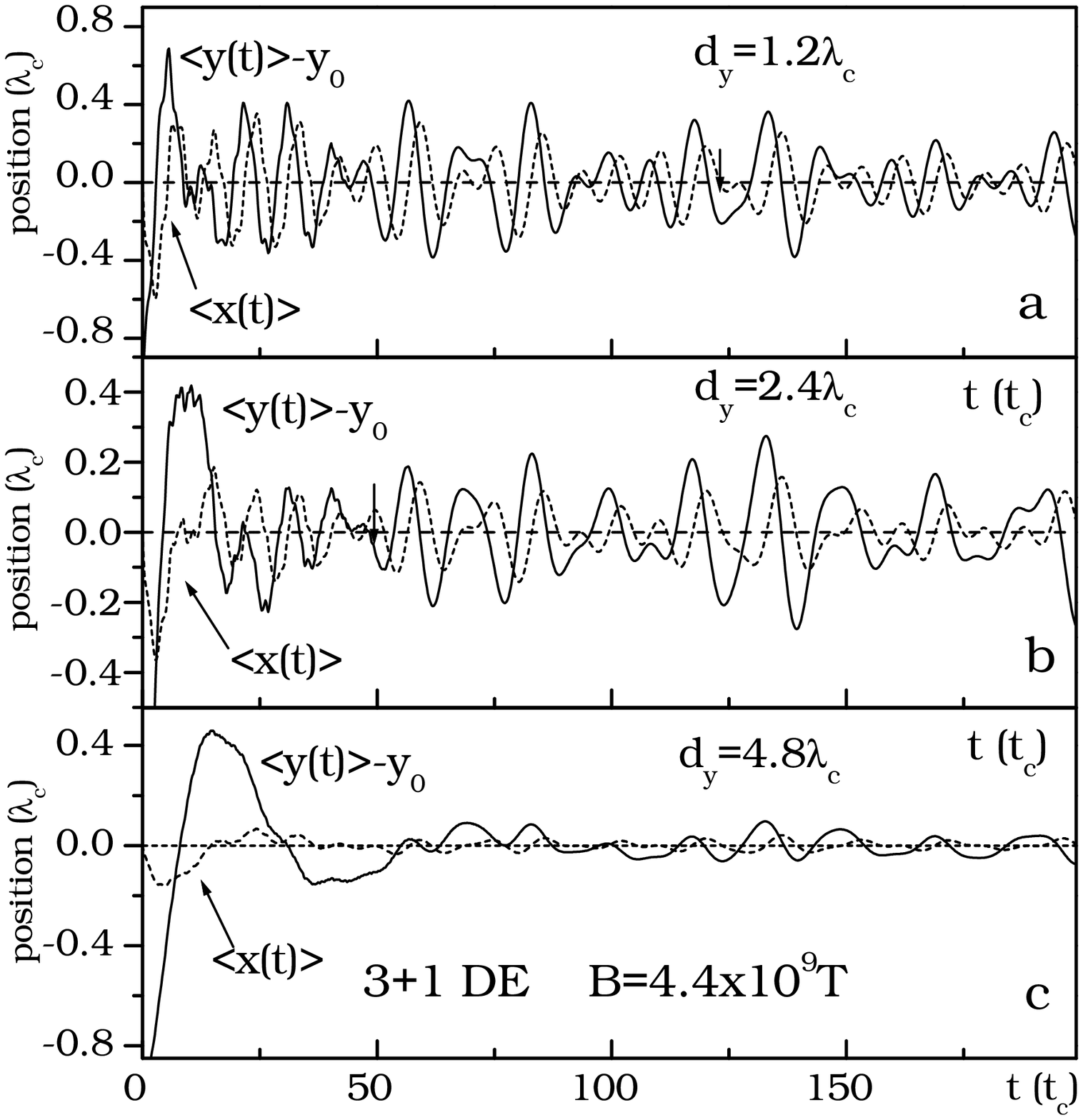}
\caption{Calculated motion of wave packet with the second nonzero component during first 200~$t_c\simeq 0.25$
         attoseconds of motion for various wave packet parameters. The magnetic field corresponds to~$L=\lambda_c$.
         Packet parameters: $d_x=1.5\lambda_c$,$d_z=1.8\lambda_c$, $k_{0x}=0.998\lambda_c^{-1}$, $k_{0z}=0$.
         Time scale is in $t_c=\hbar/(mc^2)=1.29\times$10$^{-21}$ s units, while position
         is in $\lambda_c=\hbar/(mc)=3.86\times$10$^{-13}$ m units. } \label{Fig3D1}
\end{figure}

\begin{figure}
\includegraphics[width=8.5cm,height=8.5cm]{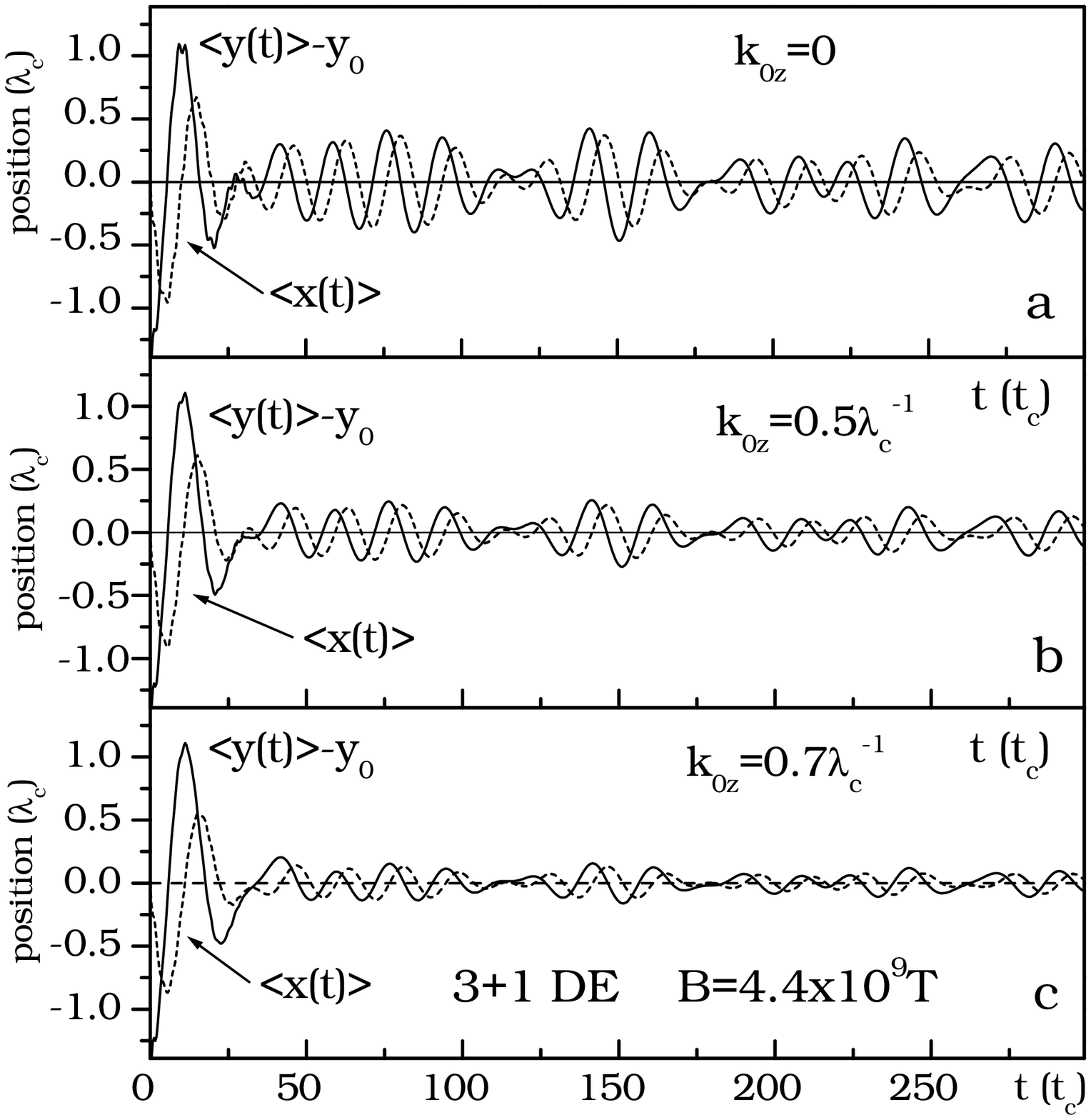}
\caption{Calculated motion of wave packet with the second nonzero component
         and nonzero velocity in the~$z$ direction.
         Packet parameters: $d_x=2.0\lambda_c$, $d_y=1.8\lambda_c$, $d_z=1.5\lambda_c$,
         $k_{0x}=0.673\lambda_c^{-1}$.} \label{Fig3D3}
\end{figure}

\begin{figure}
\includegraphics[width=8.5cm,height=8.5cm]{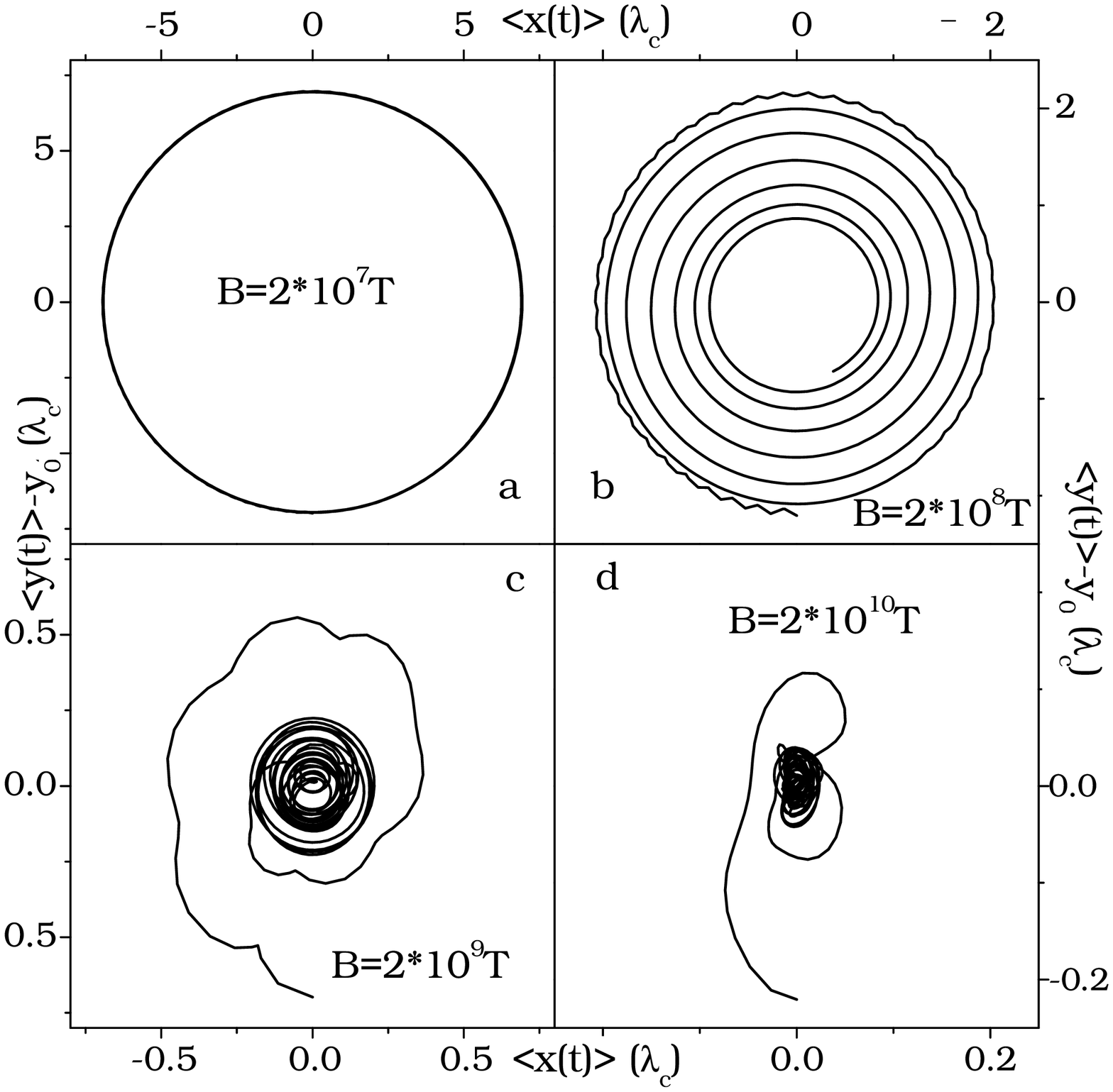}
\caption{Trajectories of wave packets with the second nonzero component for 3+1 Dirac
         equation in various magnetic fields. Packets parameters:
         $d_x=0.632(B_b/B)^{0.5}\lambda_c$,
         $d_y=0.569(B_b/B)^{0.5}\lambda_c$,
         $d_z=0.474(B_b/B)^{0.5}\lambda_c$, $k_{0z}=0$,
         $k_{0x}=0.999(B/B_b)^{0.5} \lambda_c^{-1}$, where $B_b=2\times 10^{10}$ T.
         The products $Lk_{0x}$, $d_xk_{0x}$, $d_yk_{0x}$ and $d_zk_{0x}$ are the same for all figures.
         In all cases the packet motion is transient but for lower magnetic fields
         the decay of oscillations is slow. } \label{Fig3D4}
\end{figure}

\begin{figure}
\includegraphics[width=8.5cm,height=8.5cm]{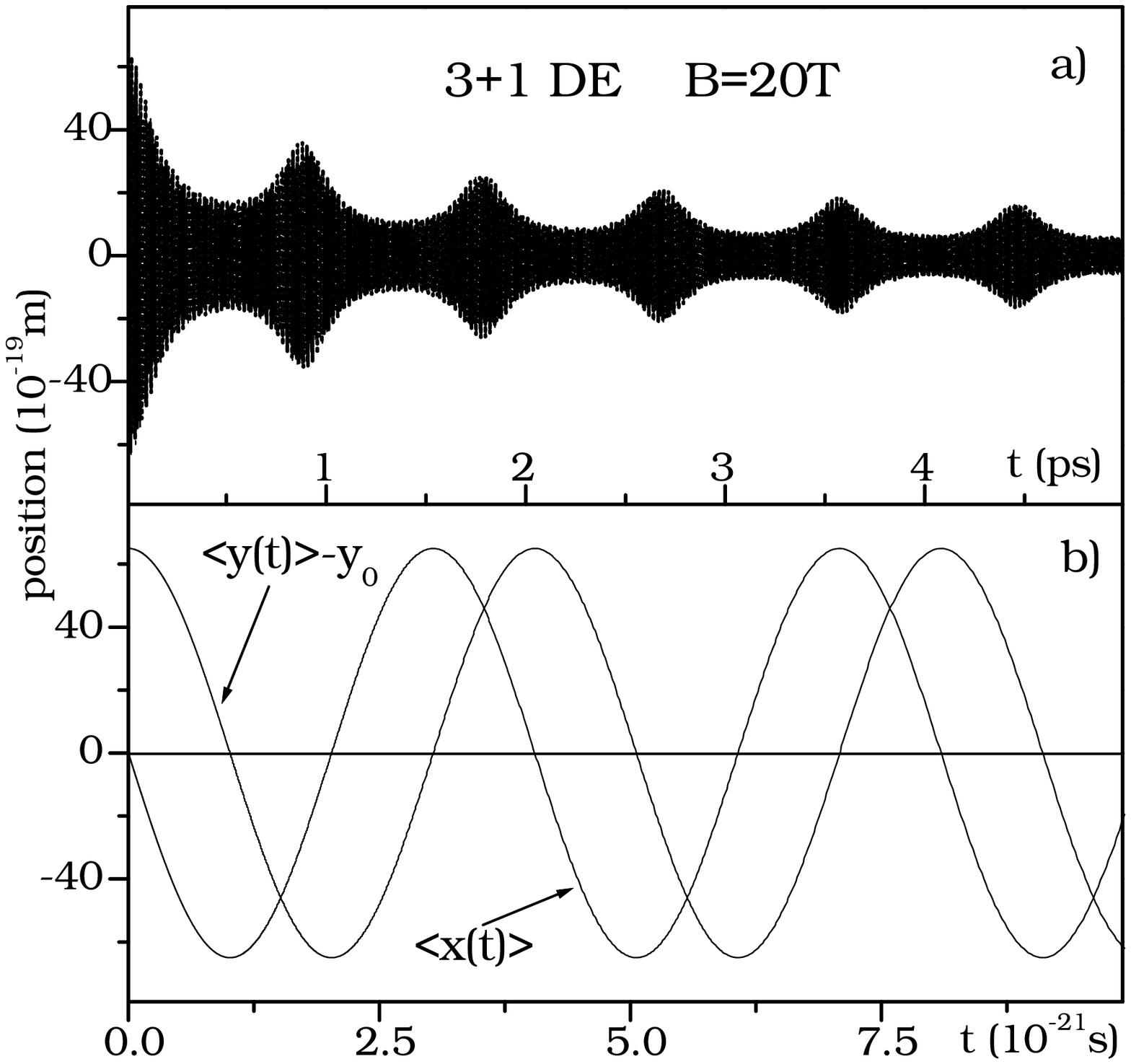}
\caption{Calculated ZB components of electron motion
         in a magnetic field in two very different time scales. Packet parameters are:
         $d_x=20000\lambda_c$, $d_y=18000\lambda_c$, $d_z=15000\lambda_c$,
         $k_{0x}=0.5L^{-1}=8.72\times$10$^7$m$^{-1}$, $k_{0z}=0$. The ZB oscillations decay as~$t^{-1/2}$
         but they have very small amplitudes.
         Note the collapse-and-revival character of ZB oscillations.} \label{Fig3DLow}
\end{figure}

We present our results for the 3+1 Dirac equation in a vacuum beginning with
the relativistic limit for a wave packet with the second nonzero component.
The average packet positions $\hY(t)$ and $\hX(t)$,
given by Eqs.~(\ref{Avg_22_Yt})-(\ref{Avg_22_Xt}), are calculated
computing numerically the coefficients~$U_{m,n}$, see Eqs.~(\ref{Gauss_Umn}) and~(\ref{Gauss_Umn_Ldy}).
In our calculations we use the first~$n=400$ Hermite polynomials.
For each set of parameters $L,d_x,d_y,k_{0x}$
we check the summation rules for~$U_{n,n}$ and $\sqrt{n+1}\ U_{n+1,n}$, see Appendix~\ref{AppendixUmn}.
With the numerical procedures we use, these rules are fulfilled with the accuracy of ten or more digits.
In Fig.~\ref{Fig3D1} we plot the electron positions calculated
for the first 200$t_c\simeq 0.25$ attoseconds of motion for various packet parameters.
The time scale is in units $t_c=\hbar/mc^2 = 1.29 \times 10^{-21}$ s.
We chose magnetic field $B=4.4\times 10^{9}$ T and an elliptic wave packet with
$k_{0x}=0.998\lambda_c^{-1}$ and~$k_{0z}=0$.
It is seen that the ZB oscillations consist of several frequencies. This is the main effect of an external
magnetic field, which quantizes both positive and negative electron energies into the Landau levels.
At larger times the oscillations in the 3+1 space go through decays and revivals, but finally disappear.
Thus the motion of electrons shown in Fig.~\ref{Fig3D1} has
a transient character in which several incommensurable frequencies
appear. The calculated motion is a combination of the intraband (cyclotron) and interband~(ZB) components.
In the relativistic regime the components have comparable amplitudes.
The character of motion, number of oscillations in the indicated time interval and
the decay times strongly depend on packet's parameters. For Fig.~\ref{Fig3D1}c we chose the packet
width $d_y=4.8\lambda_c$. The number of oscillations is then reduced compared to
Figs.~\ref{Fig3D1}a and~\ref{Fig3D1}b. This confirms a previous observation (see Ref.~\cite{Rusin20008a})
that the packet parameters have to be carefully selected for ZB to be observable.
In contrast to the low-field limit, in the high-field regime the amplitudes of ZB
are of the order of $\lambda_c$.

Motion of non-relativistic electrons in the~$z$ direction, parallel to the magnetic field,
is independent of the circular motion in~$x-y$ plane. However, the motion of
relativistic electrons in the~$z$ direction is coupled to the in-plane motion. To analyze
this effect we calculated positions of the relativistic wave packet with a nonzero initial velocity assuming
${\bm k}_0=(k_{0x},0,k_{0z})$ with the constraint $|{\bm k}_0|<\lambda_c^{-1}$.
In Fig.~\ref{Fig3D3} we show the calculated
packet motion with fixed $k_{0x}=0.673\lambda_c^{-1}$ and various values of~$k_{0z}$
for $B=4.4\times 10^{9}$ T. As seen in Figs.~\ref{Fig3D3}a,~\ref{Fig3D3}b and~\ref{Fig3D3}c, the existence
of nonzero~$k_{0z}$ component reduces the number of oscillations
in the cyclotron and ZB motions. Increasing~$k_{0z}$ leads to a faster decay of the motion.
The maximum initial amplitudes of oscillations do not depend
on~$k_{0z}$, but the amplitudes at larger times decrease with increasing~$k_{0z}$.

To visualize the gradual transition from the non-relativistic to the relativistic regime we
plot in Fig.~\ref{Fig3D4} the packet trajectories for four values of magnetic field.
In all cases the packet parameters are chosen in a systematic way keeping constant values of
the products: $Lk_{0x}=0.47$, $d_xk_{0x}=0.632$, $d_yk_{0x}=0.569$ and $d_zk_{0x}=0.474$.
For $B=2\times 10^{7}$ T the trajectories of electron motion are still circular,
as at low magnetic fields. When the field is increased to $B=2\times 10^{8}$ T the trajectories
are deformed into slowly decaying spirals.
At very high fields: $B=2\times 10^{9}$ T and $B=2\times 10^{10}$ T, the trajectories are described
by fast decaying spirals. The amplitude of motion decreases with
increasing field, which is caused by the decrease of magnetic length~$L$.

Finally, in Fig.~\ref{Fig3DLow} we plot the ZB part of motion at the low magnetic field $B=20$ T in two
scales of time. The amplitude of ZB motion is $D=6.5\times$10$^{-8}$\AA, which
agrees well with its estimation given in Eq.~(\ref{LFL_D}). In Fig.~\ref{Fig3DLow}a
we observe a slow decay of oscillations with its envelope decaying as~$t^{-1/2}$.
In Fig.~\ref{Fig3DLow}b we show rapid ZB oscillations
with the frequency $\omega_Z = 2mc^2/\hbar=7.76\times$ 10$^{20}$ s$^{-1}$.
The ZB oscillations exist even at times of the order of picoseconds.

Generally, the ZB effects are observable in magnetic fields of the order of $4.4\times 10^{9}$ T
for wave packets moving with an initial velocity close to~$c$.
These packets should have width of the order of~$\lambda_c$.
It is not possible to fulfill all these requirements using currently available experimental
techniques. In addition, the predicted amplitudes of the ZB motion are of the order of~$\lambda_c$,
which makes their experimental detection extremely difficult.
However, there exists now a very powerful experimental possibility to simulate
the Dirac equation and its consequences. We explore this possibility in the section below.

\section{Simulations by trapped ions}

\begin{figure}
\includegraphics[width=8.5cm,height=8.5cm]{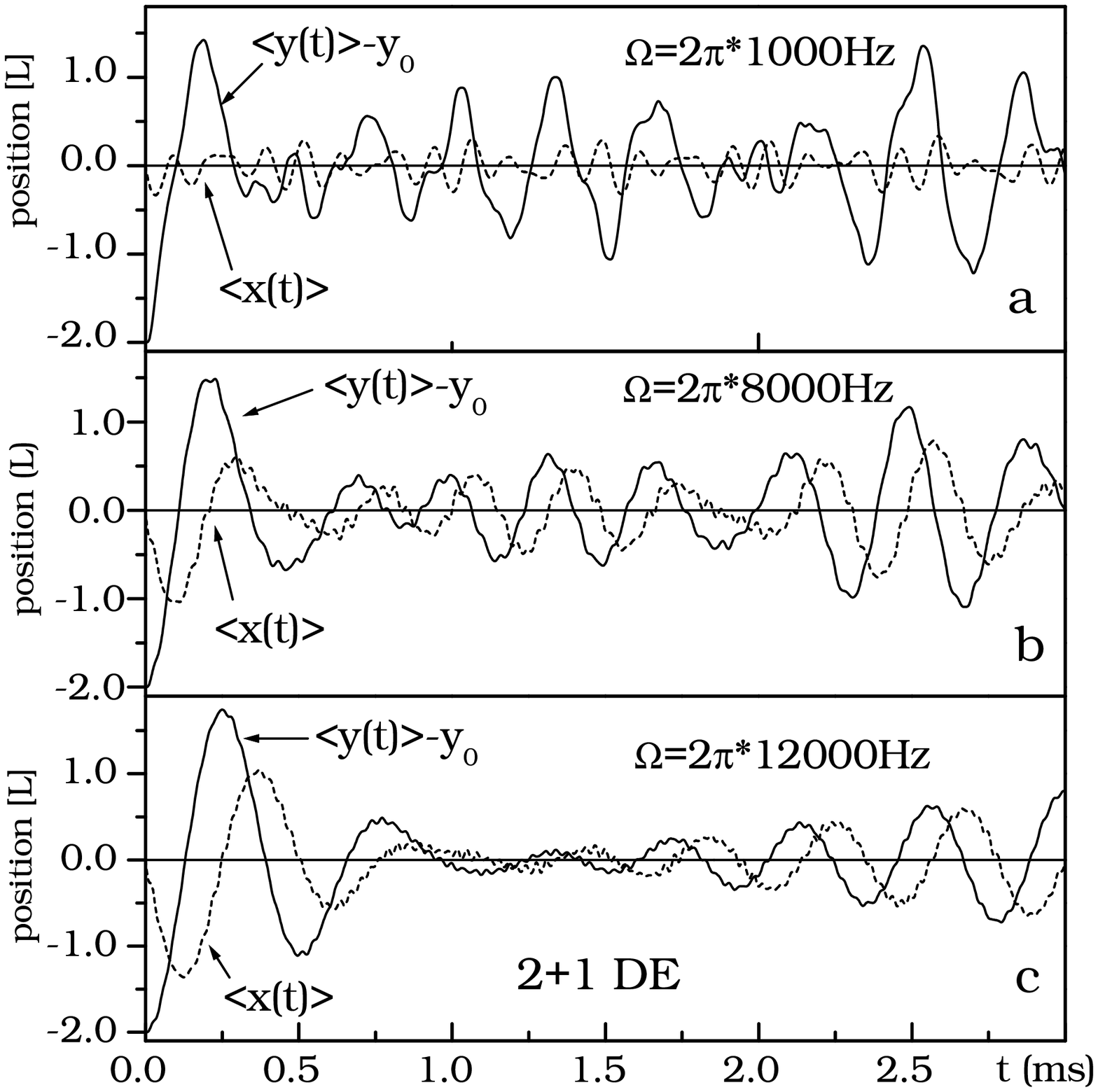}
\caption{Calculated motion of two-component wave packet simulated
         by trapped $^{40}$Ca$^{+}$ ions for three values of effective rest energies~$\hbar\Omega$.
         Trap parameters: $\eta=0.06$, $\tilde{\Omega}=2\pi\times 68$ kHz, $\Delta\simeq 96\AA$;
         packet parameters: $k_{0x}=\Delta^{-1}$, $d_y=\Delta\sqrt{2}$,
         $d_x=0.9d_y$. Simulations correspond to $\kappa=\hbar\omega_c/2mc^2$=
         16.65~(a), 0.26~(b), 0.116~(c), respectively. Positions are given in $L=\sqrt{2}\Delta$ units.
         Oscillations do not decay in time.} \label{FigSim1}
\end{figure}

\begin{figure}
\includegraphics[width=8.5cm,height=8.5cm]{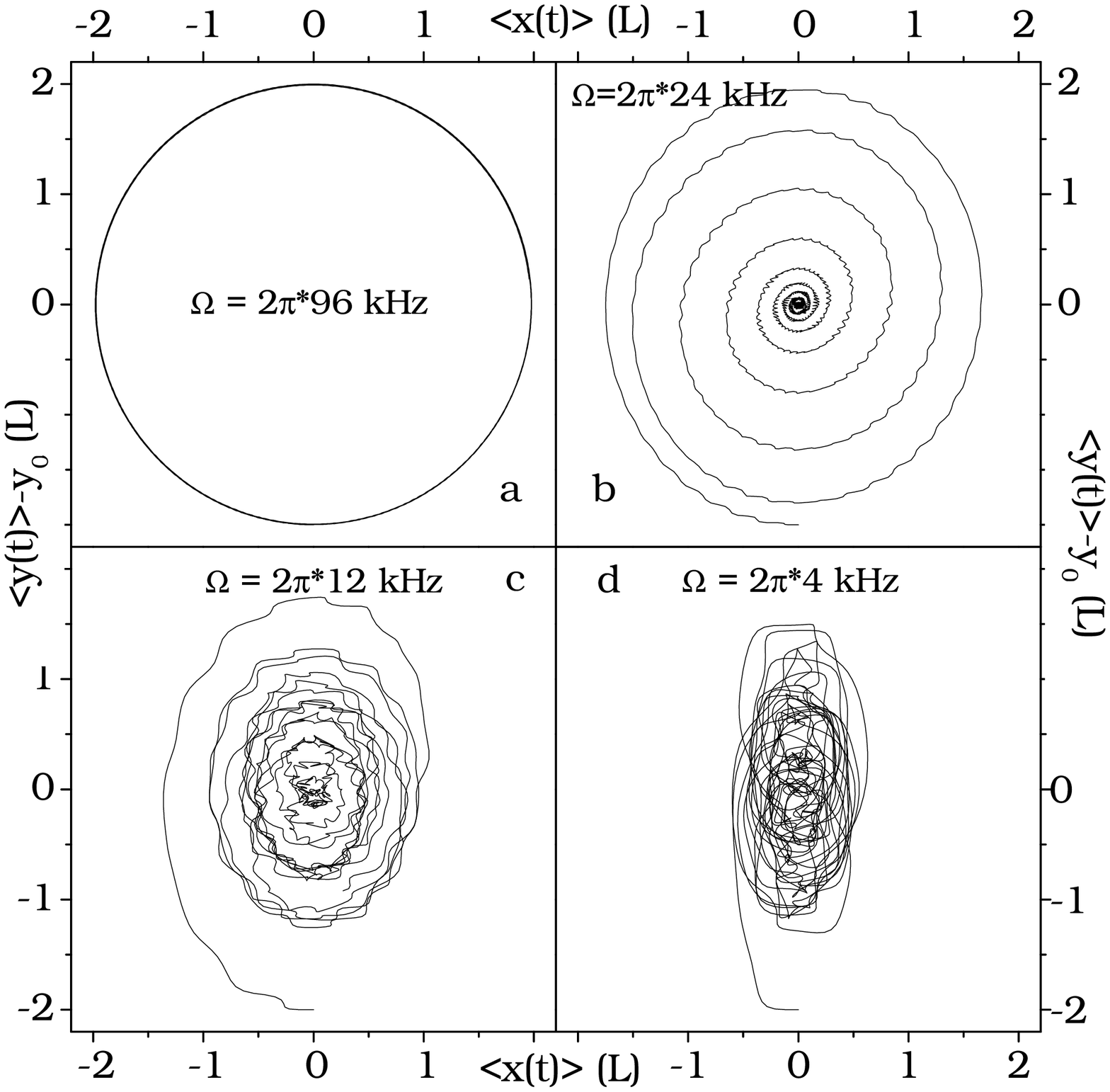}
\caption{Trajectories of electron wave packet in a constant magnetic field for various simulated rest
         energies $\hbar \Omega$, as calculated for 2+1 Dirac equation.
         Trap and packet parameters are the same as in Fig.~\ref{FigSim1}.
         Positions are given in magnetic radius~$L$.
         In the non-relativistic limit~(a) the ZB is practically absent.
         As the rest energy decreases, the motion becomes more relativistic and the ZB (interband)
         frequency components become stronger. The ratio $\kappa$ defined in Eq.~(\ref{IonB}) is:
         (a)~0.0018, (b)~0.029, (c)~0.116, (d)~1.05.}\label{FigSim2}
\end{figure}

\begin{figure}
\includegraphics[width=8.5cm,height=8.5cm]{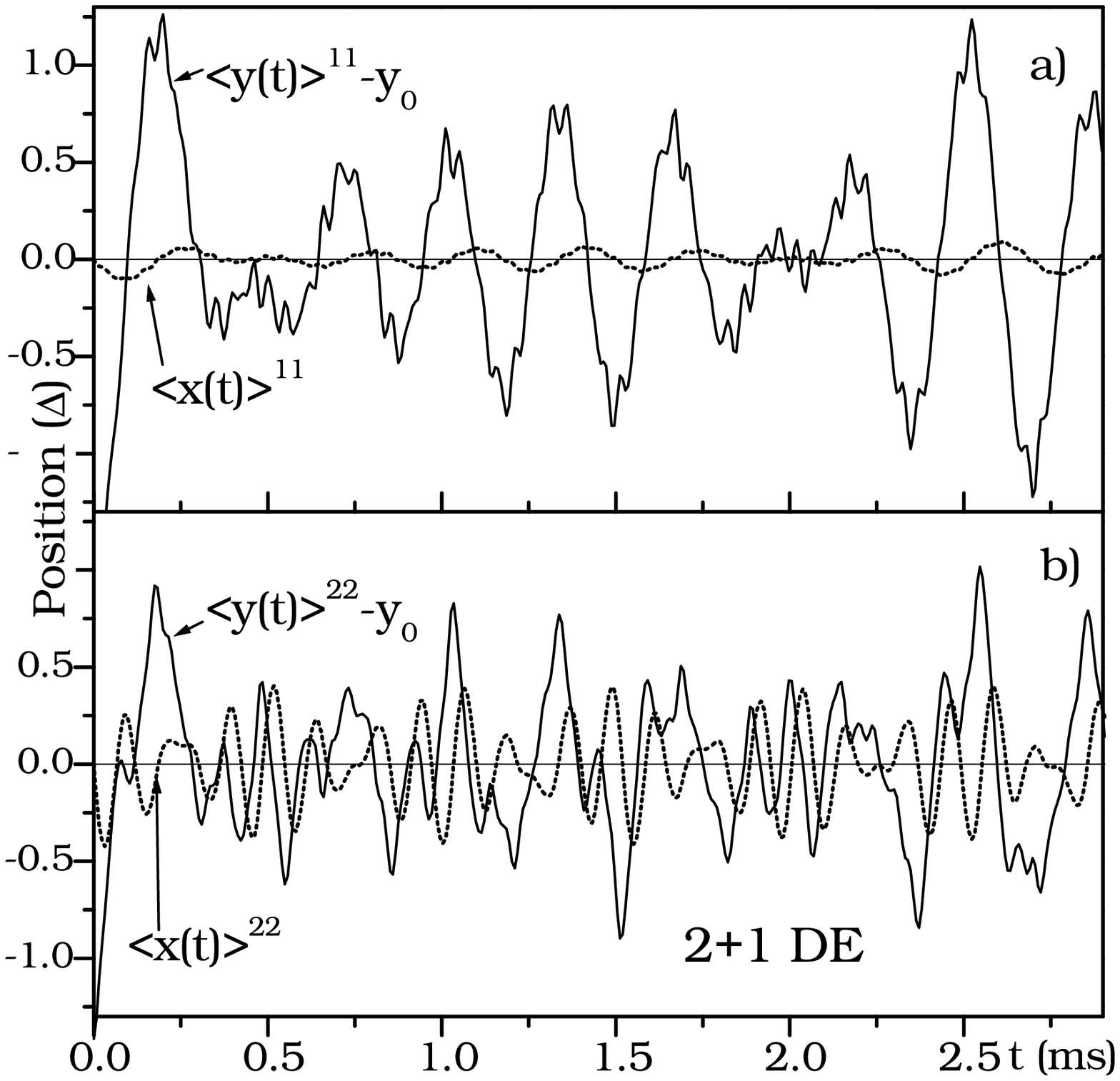}
\caption{Simulated ZB motion of one-component packets in the regime $\kappa\gg 1$.
         Simulated gap frequency is $\Omega=2\pi \times 1000$ Hz, other trap and
         packet parameters as in Fig.~\ref{FigSim1}. Upper part $-$ packet with the first
         nonzero component; lower part $-$ packet with the second nonzero component.
         Note largely different magnitudes of the~$x$ oscillations in the two cases.} \label{FigSim3}
\end{figure}

\begin{figure}
\includegraphics[width=8.5cm,height=8.5cm]{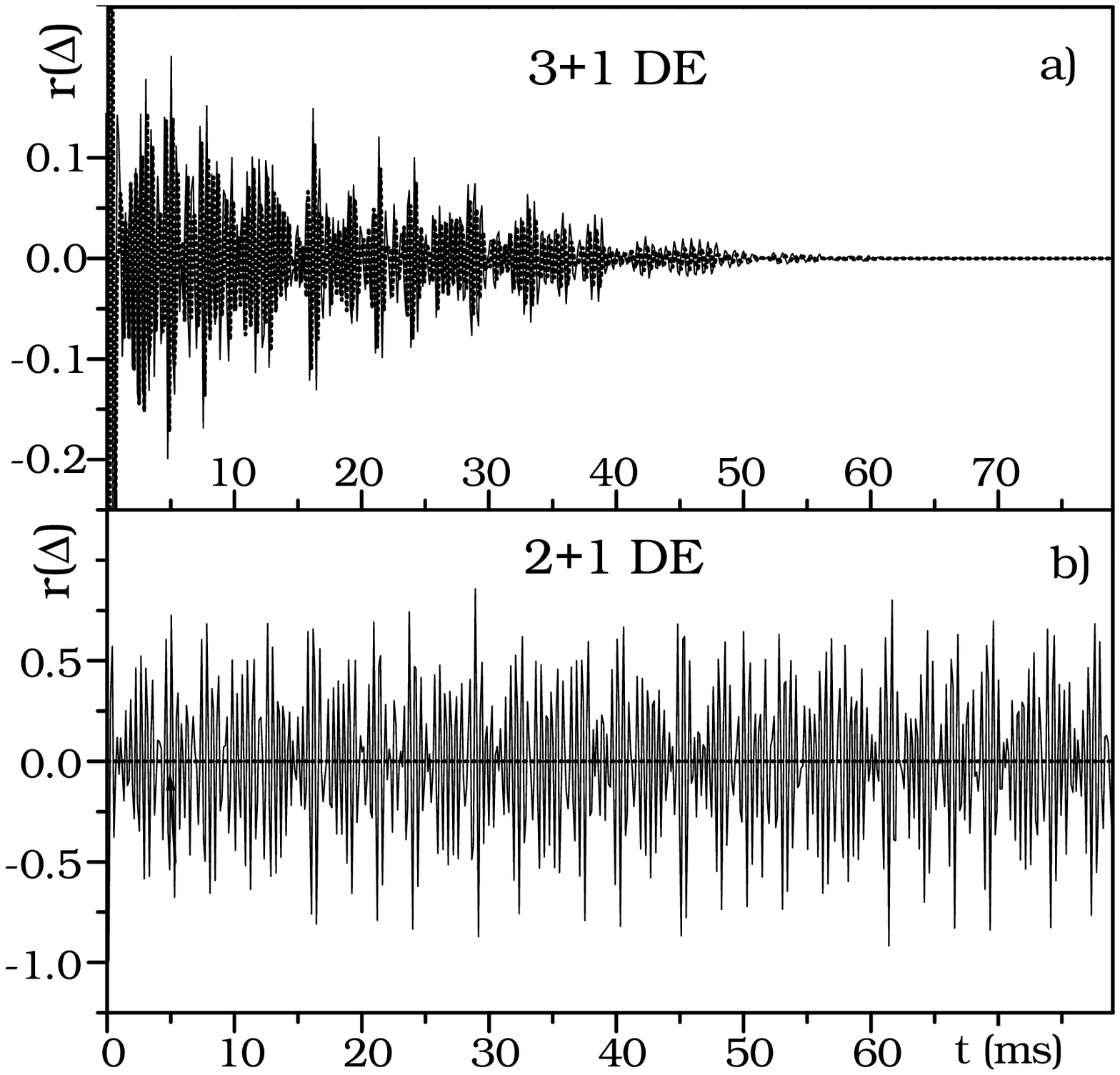}
\caption{Collapse and revivals of packet motion for simulations using 3+1 DE~(a) and 2+1 DE~(b).
         Packet parameters: $d_x=d_y=d_z=L$, $k_{0x} = \Delta^{-1}$. Trap parameters as in
         Fig.~\ref{FigSim1}, simulated gap frequency is $\Omega=2\pi\times 12000$ Hz.
         Note transient character of motion for the 3+1 DE and persistent
         oscillations for the 2+1 DE. In both cases the collapse and
         revivals appear.} \label{FigSim4}
\end{figure}

The main experimental problem in investigating the ZB phenomenon in an external magnetic field
is the fact that, for free relativistic electrons in a vacuum, the basic ZB (interband) frequency corresponds to
the energy $\hbar\omega_Z \simeq 1$~MeV, whereas the cyclotron energy for a magnetic
field of 100~T is $\hbar\omega_c \simeq 0.01$~eV. Thus the magnetic effects in ZB are
very small. However, it is now possible to simulate
the Dirac equation changing at the same time its basic parameters. This gives a possibility
to strongly modify the critical ratio $\hbar\omega_c/2mc^2$ making it more advantageous.
In the following we propose how to simulate the 3+1 and 2+1 Dirac equations in the presence of a
magnetic field using trapped ions and laser excitations.

First, we transform the Dirac equation to the off-diagonal form
\begin{equation}\hHD^{'}= c\sum_{i}\alpha_i\hat{p}_i+\delta mc^2, \end{equation}
using the unitary operator
$\hat{P}=\delta(\delta+\beta)/\sqrt{2}$, where
$\delta=\alpha_x\alpha_y\alpha_z\beta$~\cite{Moss1976}. After the transformation the Hamiltonian is
$\hHD^{'}=\left(\begin{array}{cc} 0 &\hH^{'} \\ \hH^{'\dagger} &0 \end{array} \right)$,
where
\begin{equation} \label{IonHD}
\hH^{'} = \left(\begin{array}{cc} c\hat{p_z} - imc^2 &c\hat{p_x}-\hbar\omega \ha_y \\
     c\hat{p_x}-\hbar\omega \hap_y &-c\hat{p_z} - imc^2 \end{array} \right),
\end{equation}
and~$\ha_y$ and~$\hap_y$ are given in Eq.~(\ref{H_aap_def}).

Next we use the procedures developed earlier
and consider a four-level system of Ca or Mg trapped ions~\cite{Lamata2007,Johanning2009,Leibfried2003}.
Simulations of~$cp_x$ and~$cp_z$
terms in the above Hamiltonian are carried out the same way as for free Dirac particles using
pairs of the Jaynes-Cumminngs~(JC) interactions
\begin{equation}
\hH^{\phi_r}_{JC}= \hbar\eta\tilde{\Omega}(\sigma^+\ha e^{i\phi_r}+\sigma^-\hap e^{-i\phi_r}),
\end{equation}
and the anti-Jaynes-Cumminngs~(AJC) interactions
\begin{equation}
\hH^{\phi_b}_{AJC}= \hbar\eta\tilde{\Omega}(\sigma^+\hap e^{i\phi_b}+\sigma^-\ha e^{-i\phi_b}).
\end{equation}
A simulation of~$mc^2$ is done by the so called carrier interaction
\begin{equation}\hH^{c}= \hbar\Omega(\sigma^+e^{i\phi_c}+\sigma^-e^{-i\phi_c}).\end{equation}
Here~$\Omega$ and~$\tilde{\Omega}$ are coupling strengths and~$\eta$ is the Lamb-Dicke parameter~\cite{Lamata2007}.
The operators~$\ha$ and~$\hap$ are lowering and raising operators of the one-dimensional
harmonic oscillator, respectively.
These operators can be associated with the three normal trap frequencies and, therefore, with the motion along
the three trap axes. Setting pairs of lasers beams in the~$x$,~$y$ and~$z$ directions
it is possible to simulate the lowering and raising operators along these directions, respectively.
As an example of this procedure, one selects a pair of JC and AJC interactions in the~$x$ direction,
adjusting their phases $\phi_r=-\pi/2$ and $\phi_b=+\pi/2$. This way one can simulate the $2\times 2$ Hamiltonian
$\hH^{p_x}_{\sigma_x} = \hH^{\phi_r}_{JC} + \hH^{\phi_r}_{JC}$ to get
\begin{equation}
\hH^{p_x}_{\sigma_x} = i\hbar\eta\tilde{\Omega} \sigma_x (\hap_x-\hap_x)
   = 2\eta_q\tilde{\Omega}\sigma_x \Delta_x p_x,
\end{equation}
where $p_x = i\hbar(\hap_x-\ha_x)/\Delta_x$. Using this technique, the~$p_x$ and~$p_z$ dependent
parts of the Dirac Hamiltonian~(\ref{IonHD}) can be simulated by appropriate
combinations of JC and AJC interactions.
On the other hand, a simulation of~$\ha_y$ and~$\hap_y$ operators (which include the magnetic field)
can be done by single JC or AJC interactions. Using the notation of
Refs.~\cite{Lamata2007,Johanning2009,Leibfried2003}
one simulates the complete 3+1 Hamiltonian~$\hHD^{'}$ by the following set of excitations
\begin{eqnarray}  \label{IonSim}
\hHD^{'}_{ion} &=& \hH^{p_x}_{\sigma_x (ad)}+ \hH^{p_x}_{\sigma_x (bc)}+
               \hH^{\phi_r=\pi}_{JC (ad)}  + \hH^{\phi_b=\pi}_{AJC (bc)}  + \nonumber \\
&&             \hH^{p_z}_{\sigma_x (ac)}- \hH^{p_z}_{\sigma_x (bd)}+
               \hH^{c}_{\sigma_y (ac)}+ \hH^{c}_{\sigma_y (bd)},
\end{eqnarray}
where $\hH^{p_q}_{\sigma_j}= 2\eta_q\tilde{\Omega}\sigma_j \Delta_q p_q$, $p_q = i\hbar(\hap_q-\ha_q)/\Delta_q$,
$j,q=x,z$. The subscripts in parentheses of Eq.~(\ref{IonSim})
symbolize states involved in the transition in question.
The spread of the ground ion wave function is $\Delta_q=\sqrt{\hbar/2M\nu_q}$
and the Lamb-Dicke parameter is $\eta_q=k\sqrt{\hbar/2M\nu_q}$, where
$M$ is ion's mass,~$\nu_q$ is trap's frequency in the~${\bf q}$ direction
and~${\bf k}$ is the wave vector of the driving field in a trap.
The JC interaction gives~$\ha_y$ in~$\hH^{'}_{12}$ and~$\hap_y$ in $\hH^{'\dagger}_{21}$
elements of the Hamiltonian~$\hH^{'}$ in Eq.~(\ref{IonHD}), respectively,
while AJC gives~$\ha_y$ in $\hH^{'}_{21}$ and $\hap_y$ in $\hH^{'\dagger}_{12}$ elements, respectively.
A simulation of the 3+1 DE by Eq.~(\ref{IonSim}) can be realized with 12 pairs of laser excitations:
two pairs for each of the four interactions simulating~$p_x$ and~$p_z$ terms and
one pair for each of the four remaining terms.
If one omits the~$p_z$ interaction, which corresponds to the 2+1 DE, one needs 8 pairs of laser excitations:
two pairs for the~$p_x$ terms and one pair for the each of four remaining terms.
Simulated magnetic field intensity can be found from the following correspondence (see Ref.~\cite{Lamata2007}):
$\ha_y-\hap_y=\sqrt{2}L(\partial/\partial y)= 2\Delta (\partial/\partial y)$, which gives
$L/\sqrt{2} \Leftrightarrow \Delta $, where $\Delta_x=\Delta_y=\Delta_z=\Delta$.
Since the other simulations are: $c \Leftrightarrow 2\eta\Delta\tilde{\Omega}$ and
$mc^2 \Leftrightarrow \hbar \Omega$, we have for the critical ratio
\begin{equation}  \label{IonB}
 \kappa = \frac{\hbar eB}{m(2mc^2)} \Leftrightarrow \left(\frac{\eta\tilde{\Omega}}{\Omega}\right)^2.
\end{equation}
Therefore, by adjusting the frequencies ~$\Omega$ and~$\tilde{\Omega}$ one simulates different values of
$\kappa=\hbar\omega_c/2mc^2$. This illustrates the fundamental advantage of simulations by trapped ions.

In Fig.~\ref{FigSim1} we show the calculated Zitterbewegung for three values of~$\kappa$:
16.65, 1.05, 0.116, using a two-component electron wave packet
$\langle {\bm r}|f\rangle =f({\bm r})(\sqrt{2}/2,\sqrt{2}/2,0,0)$.
The electron motion is a combination of
$\langle \hY \rangle^{1,1}(t)$,    $\langle \hY \rangle^{2,2}(t)$,
$\langle \hY \rangle^{1,1}(t)$ and $\langle \hX \rangle^{2,2}(t)$ components.
There are no mixing terms of the form  $\langle \hY\rangle^{1,2}(t)$ etc., since
they vanish for the 2+1 Dirac equation due to their proportionality to~$p_z$, see Eq.~(\ref{Avg_J}).
The essential feature of the simulated characteristics is their low frequency and
large amplitude of ZB. Further, it is seen that, as~$\kappa$ gets larger
(i.e. the field intensity increases or the
effective gap decreases), the frequency spectrum of ZB becomes richer.
This means that more interband and intraband frequencies contribute to the spectrum.
Both types of frequencies correspond to the selection rules~$n'=n\pm 1$.
Thus, for example, one deals with ZB (interband) energies between the Landau levels~$n=0$ and~$n'=1$,
and~$n=1$ and~$n'=0$, as the strongest contributions. For simulated high magnetic fields
corresponding to~$\kappa \ge 1$, the interband and intraband
components are comparable and one can legitimately talk about ZB.
We believe that the ZB oscillations of the type shown in Fig.~\ref{FigSim1}a,
resulting from the 2+1 DE for
$\kappa=\hbar\omega_c/2mc^2 > 1$, are the best candidate for an observation of the simulated trembling
motion in the presence of a magnetic field. The calculated spectra use the
trap and wave packet parameters already realized experimentally, see~\cite{Gerritsma2010}.
We emphasize the tremendous differences of the position scales between the results
for free electrons in a vacuum, shown in Fig.~\ref{Fig3D1}, and the simulated ones shown in Fig.~\ref{FigSim1}.
The anisotropy of ZB with respect to $\langle x(t)\rangle$
and $\langle y(t)\rangle$ components, seen in Figs.~\ref{Fig3D1} and~\ref{FigSim1},
is due to the initial conditions, namely~$k_{0x}\neq 0$
and~$k_{0y}=0$. A similar anisotropy was predicted in the zero-gap
situation in graphene~\cite{Rusin20008a}.

In Fig.~\ref{FigSim2} we show the results of our calculations
for different~$\hbar\Omega$, simulating effective values of~$2mc^2$,
at a constant simulated magnetic field. Packet parameters are the same as in Fig.~\ref{FigSim1}.
The results are shown for initial time intervals of the motion.
In the non-relativistic limit illustrated in Fig.~\ref{FigSim2}a, the motion is completely
dominated by the intraband frequencies and it represents a cyclotron orbit. As the gap decreases,
the motion is more relativistic and the circular trajectories
turn into spirals. Simultaneously, the interband Zitterbewegung frequencies come into play.
In highly relativistic regimes (low values of~$\hbar \Omega$)
the trajectories look chaotically. However, the motion is not chaotic, it consists of a finite
number of well defined but incommensurable frequencies.
The illustrated motion of the wave packet for the 2+1 Dirac equation is persistent, its amplitude
experiences infinite series of collapse and revival cycles.
In the relativistic regime the motion is somewhat anisotropic with respect to~the $x$ and~$y$
directions which is related to the initial conditions ${\bm k}_0=(k_{0x},0)$.
This phenomenon has an analogy in the field-free case for the
relativistic regime, where the ZB oscillations occur in the direction perpendicular to the
initial packet velocity~\cite{Schliemann05,Rusin07b}.

In Fig.~\ref{FigSim3} we analyze ZB of the one-component packets having a non-vanishing first or
second component. Interestingly, they look distinctly different, and the~$x$ parts of the motion
have different limits for~$mc^2\rightarrow 0$ (i.e. for very small energy gaps~$\hbar\Omega$).
The~$y$ components of motion are comparable in both cases, but the~$x$ components differ substantially.

In all the figures presented above we showed the packet motion in short time spans.
In Fig.~\ref{FigSim4} we analyze the long-time packet evolution according to the simulated 3+1 and 2+1
Dirac equations. In both cases the collapse and revival cycles occur.
However, the motion according to the 3+1 Dirac equation is decaying in time,
while the oscillations in the 2+1 case are persistent in time.

\section{Discussion}

We briefly summarize the important new effects brought to ZB by an external magnetic field:
(1) The quantization of the spectrum for positive and negative electron energies results
in numerous interband frequencies contributing to ZB,
(2) The presence of~$B$ introduces an important new parameter into the phenomenon of ZB
affecting all the frequencies,
(3) The presence of intraband frequencies raises the question of what should be and what should not be called ZB.
In our opinion, the interband frequencies are the signature of ZB while the intraband frequencies
(the cyclotron resonance in our case) are not,
(4) The presence of~$B$ 'stabilizes' ZB in the 2+1 case making it a stationary phenomenon, not decaying in time.
The last feature is related to the fact that the magnetic field is represented by a quadratic
potential and, as is well known, the wave packet in a parabolic
potential is not spreading in time. However, a slow decay of ZB in time
might occur if the trembling electron emits radiation. This does not occur if the electron
is in its eigenstate but it will happen if the electron is prepared in the form of a wave packet,
because the latter contains numerous eigenstates of the
electron in a magnetic field, see Eq.~(\ref{M_Lippman}). The emitted radiation can have multipole character
depending on the electron energy \cite{Ginzburg1965,Ginzburg1969,Erber1966},
it may also be due to spontaneous radiative
transitions between various Landau levels in the strictly quantum limit.
Finally, in the classical limit of very high electron energies one may deal with the synchrotron radiation,
radiative damping, etc., but this limit is beyond the scope of our paper.
Also, a broadening of Landau levels due to external
perturbations results in a transient character of ZB, c.f.~\cite{Rusin2009}.

The time-dependent electron motion, as obtained in the operator form [see Eqs.~(\ref{H_Yt})-(\ref{H_Xt})],
is described by four operators. We show in Appendix~\ref{Appendix4Packets} that these operators have
different limits for low magnetic fields. However, all of them contain both interband and intraband
frequencies. Thus, in both operator and average formulations the cyclotron and trembling motion
components are mixed. The method of direct averaging of operators in the Heisenberg form,
used in Section III, is simpler than that of averaging the explicit
forms of~${\hAD}$ and~${\hApD}$, as derived in Section II, since it does not require the
detailed knowledge of these operators. The main disadvantage of the direct averaging is
that it obscures the detailed structure of electron motion shown in Eqs.~(\ref{H_Yt})-(\ref{H_Xt}).

In our considerations we used one-component and two-component wave packets
and showed that the character of ZB oscillations in the two cases is
similar, but not identical. Calculation for three- and four-component packets, although
possible, are much more complicated not introducing anything new at the
physical level.

High magnetic fields for relativistic electrons in a vacuum are often characterized
by the so-called Schwinger critical field $B_{cr}$ for which $\hbar eB/m = mc^2$ or,
equivalently, $L=(\hbar/eB)^{1/2}= \hbar/mc=\lambda_c$. This corresponds to the gigantic
field $B_{cr}=4.4\times 10^9$ T existing only in the vicinity of neutron stars. However,
in simulating the analogous situations in semiconductors~\cite{Zawadzki2005KP}
or by trapped ions~\cite{Gerritsma2010}, the corresponding critical fields are not high
and they depend on parameters of the system in question. We emphasize that
our results are not limited by any particular value of $B$ and they describe
both weak and high field limits.

As mentioned in the Introduction, the initial Dirac equation~(\ref{H_DE}) and our
resulting calculations, as well as the simulations based on trapped ions,
represent the 'empty' Dirac Hamiltonian which does not take into account
the 'Fermi sea' of electrons in a vacuum having negative energies. This
one-electron model follows the original considerations of Schrodinger's.
The phenomenon of electron ZB in a vacuum is commonly interpreted as
resulting from an interference of electron states corresponding to
positive and negative electron energies. The characteristic interband
frequency of ZB is a direct consequence of this feature. The initial
electron wave packet must contain these positive and negative energy
components. It may be difficult to prepare such a packet if all negative
energies are occupied. What is more, the fully occupied negative energies
may prevent the interference (and hence ZB) to occur, see~\cite{Krekora2004,Barut1968}.
It has been a matter of controversy what happens when an electron-positron hole
pair is created by a gamma quantum~\cite{Wang2008}. On the other hand, a system with
negative electron energies can be relatively easily created in
semiconductors, see~\cite{Zawadzki2005KP}. It should be mentioned that an external magnetic
field does not create by itself the electron-positron pairs. We emphasize
again that our present calculations and the experimental simulation of
Ref.~\cite{Gerritsma2010} are realized for the one-electron Dirac equation for which ZB
certainly exists.

Bermudez {\it et al.}~\cite{Bermudez2007} treated the problem of time dependent relativistic
Landau states by mapping the relativistic model of electrons in a magnetic
field onto a combination of the Jaynes-Cummings and Anti-Jaynes-Cummings
interactions known from quantum optics. For simplicity the $p_z=0$
restriction was assumed. Three regimes of high (macroscopic), small
(microscopic) and intermediate (mesoscopic) Landau quantum numbers n were
considered. In all the cases one interband frequency contributed to the
Zitterbewegung because the authors did not use a gaussian wave packet to
calculate average values.

Our exact calculation of Zitterbewegung of relativistic electrons in a
vacuum in the presence of a magnetic field and its simulation by
trapped ions are in close relation with the proof-of-principle
experiment of Gerritsma {\it et al.}~\cite{Gerritsma2010}, who simulated the 1+1 Dirac equation
and the resulting electron ZB in absence of magnetic field. Our results
show that, paradoxically, the simulation of the DE with a magnetic
field is simpler than that without the field. However, there is a price
to pay: one needs at least the 2+1 DE to describe the magnetic motion
since ${\bf B}$ parallel~$z$ couples the electron motion in~$x$ and~$y$ directions.

\section{Summary}

In summary, we treated the problem of electron Zitterbewegung in the
presence of a magnetic field in three ways. First, we carried
calculations at the operator level deriving from the one-electron Dirac
equation the exact and analytical time-dependent equations of motion for
appropriate operators and finally for the electron trajectory. It
turned out that, in the presence of a magnetic field, the electron motion
contains both intraband and interband frequency components, which we
identified as the cyclotron motion and the trembling motion~(ZB),
respectively. Next, we described the same problem using averages of the
Heisenberg time-dependent operators over Gaussian wave packets in order to
obtain physical quantities directly comparable to possible experimental
verifications. We found that, in addition to the usual problems with the
very high frequency and very small amplitude of electron Zitterbewegung in
a vacuum, the effects of a magnetic field achievable in terrestrial
conditions on ZB are very small. In view of this, we simulated the Dirac
equation with the use of trapped atomic ions and laser excitations in
order to achieve more favorable ratios of $(\hbar eB/m)/(2mc^2)$ than those
achievable in a vacuum, in the spirit of recently realized experimental
simulations of the 1+1 Dirac equation and the resulting electron
Zitterbewegung. Various characteristics of the relativistic electron
motion were investigated and we found that the influence of a simulated
magnetic field on ZB is considerable and certainly observable.
It was shown that the 3+1 Dirac equation describes decaying ZB oscillations
while the 2+1 Dirac equation describes stationary ZB oscillations.
We hope that our theoretical predictions will prompt experimental simulations of
electron Zitterbewegung in the presence of a magnetic field.

\appendix

\section{} \label{AppendixXY}
In this Appendix we briefly summarize the similarities and differences between operators $\hY$ and $\hX$, as
defined in Eqs.~(\ref{H_Y})-(\ref{H_X}), and the position operators $\hat{y}$ and $\hat{x}$. The operators
$\hY = (L/\sqrt{2})(\ha+\hap){\rm diag}(1,1,1,1)$ and $\hX=(L/i\sqrt{2})(\ha-\hap){\rm diag}(1,1,1,1)$
are $4\times 4$ non-commuting matrices: $[\hX,\hY]=1$, while the position operators~$\hat{y}$,~$\hat{x}$
obviously commute. However, the matrix elements of
$\hY$ and $\hX$ between states $|{\rm n}\rangle$ and $|{\rm n'}\rangle$,
given in Eq.~(\ref{M_Lippman}), are equal (up to a constant $y_0=k_xL^2$), to the matrix
elements of $\hat{y}$, $\hat{x}$ between the same states.

As an example of this property we calculate the matrix elements of $\hY$, $\hX$,
$\hat{y}$, $\hat{x}$ at $t=0$ between two states $|{\rm n}\rangle =|n,k_x,k_z,\epsilon,-1\rangle$ and
$|{\rm n'}\rangle =|n',k_x',k_z',\epsilon',-1\rangle$ given in Eq.~(\ref{M_Lippman}).
We have
\begin{eqnarray} \label{AppXY_Y}
\langle {\rm n}|\hY| {\rm n'}\rangle &=&
\frac{L}{\sqrt{2}} \left\{\langle n|\ha+\hap| n'\rangle  (\chi_{n}\chi_{n'} + N_{n} N_{n'}c^2p_z^2)
 \right. \nonumber\\
      &+& \left.\langle n-1|\ha+\hap| n'-1\rangle N _{n} N_{n'}\hbar^2 \omega_n\omega_{n'}\right\}, \ \ \ \ \ \
\end{eqnarray}
where $|n\rangle$ is defined in Eq.~(\ref{M_nkxkz}) and we omitted indices $k_x$ and $k_z$.
For the matrix element $\langle {\rm n}|\hat{y}| {\rm n'}\rangle$ we obtain the same
expression as in Eq.~(\ref{AppXY_Y})
but with $(L/\sqrt{2})(\ha+\hap)$ replaced by $\hat{y}$.
Because $(L/\sqrt{2})(\ha+\hap)= \hat{y} - k_xL^2$ we obtain from Eq.~(\ref{AppXY_Y})
\begin{eqnarray} \label{AppXY_Ywyn}
\langle {\rm n}|\hY| {\rm n'}\rangle &=& \langle {\rm n}|\hat{y}| {\rm n'}\rangle -
\langle n|k_xL^2| n'\rangle  (\chi_{n}\chi_{n'} + N_{n} N_{n'}c^2p_z^2)  \nonumber\\
      &-& \langle n-1|k_xL^2| n'-1\rangle N _{n} N_{n'} \hbar^2\omega_n\omega_{n'} \nonumber\\
     &=& \langle {\rm n}|\hat{y}| {\rm n'}\rangle - k_xL^2.
\end{eqnarray}

In order to calculate the matrix elements of $\hat{x}$ we observe that the Hamilton equations give:
$\dot{\hat{x}} = c\hat{\alpha}_x$, $\dot{\hat{y}} = c\hat{\alpha}_y$, $\dot{\hat{p}}_x=0$ and
$\dot{\hat{p}}_y=c\hat{\alpha}_x eB$. From the above relations one obtains
$\dot{\hat{p}}_y = eB \dot{\hat{x}}=(\hbar/L^2)\dot{\hat{x}}$,
which gives after the integration over time
\begin{equation}
 \hat{x}(t) = (L^2/\hbar) \hat{p}_y(t) + D.
\end{equation}
The constant of integration $D$ can be set equal to zero by an appropriate choice of $\hat{x}(0)$.
Since $\hat{p}_y = (\hbar/i)\partial/\partial y$ with $\partial/\partial y = (1/L)\partial/\partial \xi$
and $\partial/\partial \xi=(\ha-\hap)/\sqrt{2}$ [see Eq.~(\ref{H_aap_def})], there is
$\hat{p}_y(t)=  (\hbar/iL\sqrt{2})(\hat{\cal A}(t)-\hat{\cal A}^{\dagger}(t))$,
see Eqs.~(\ref{H_hAD_def})-(\ref{H_hApD_def}).
Thus we have
\begin{equation} \label{AppXY_Xwyn}
 \langle {\rm n}|\hat{x}(t)|{\rm n'}\rangle =
  \langle {\rm n}|\frac{L}{i\sqrt{2}}(\hat{\cal A}(t)-\hat{\cal A}^{\dagger}(t))|{\rm n'}\rangle =
   \langle {\rm n}|\hat{\cal X}(t)|{\rm n'}\rangle.
\end{equation}
Since $\hat{\cal A}$ and $\hat{\cal A}^{\dagger}$ are four-component lowering and raising operators, the
selection rules for $\hat{x}$ and for $\hat{\cal X}$ are $n'=n\pm 1$, $k_x=k_x'$ and $k_z=k_z'$.
There is no selection rules for $\epsilon,\epsilon'$ and for $s,s'$.
Equations~(\ref{AppXY_Ywyn}) and~(\ref{AppXY_Xwyn}) are the required relations between
the matrix elements of $\hY$, $\hX$ and $\hat{y}$, $\hat{x}$ operators, respectively.

For the states $|{\rm n}\rangle$ and $|{\rm n'}\rangle$ with $s=+1$ there is also
$\langle {\rm n}|\hY| {\rm n'}\rangle = \langle {\rm n}|\hat{y}| {\rm n'}\rangle - k_xL^2$ and
$\langle {\rm n}|\hX| {\rm n'}\rangle = \langle {\rm n}|\hat{x}| {\rm n'}\rangle$. For the states
$|{\rm n}\rangle$ and $|{\rm n'}\rangle$ with different spin indexes $s$ and $s'$ the constant term
$y_0=k_xL^2$ does not appear.

Finally we calculate the average values of $\hat{y}$, $\hat{x}$, $\hY$ and $\hX$ operators
using a Gaussian wave packet $|f\rangle$ from Eq.~(\ref{Gauss_packet}). At $t=0$
there is $\langle f|\hat{y}|f\rangle=0$ and $\langle f|\hat{x}|f\rangle=0$. Next,
\begin{equation}
\langle f|\hX|f\rangle = L\langle f|\frac{\partial}{\partial \xi}|f\rangle =
L\frac{\partial y}{\partial\xi} \langle f|\frac{\partial}{\partial y}|f\rangle = 0,
\end{equation}
and
\begin{equation}
\langle f|\hY|f\rangle = \langle f|\hat{y}|f\rangle - \langle f|k_xL^2|f\rangle = -k_{0x}L^2.
\end{equation}
All figures above refer to the averages $\langle \hY(t) \rangle$ and $\langle \hX(t) \rangle$
i.e., equivalently, to $\langle \hat{y}(t) \rangle-y_0$, $\langle\hat{x}(t)\rangle$,
respectively.

\section{} \label{AppendixCheck}

\begin{widetext}

\begin{table}[t]
 \begin{tabular}{|r|c|c|c|c|}
  \hline
 Operator & $(+1,+1)$ & (+1,-1) & (-1,+1) & (-1,-1) \\
 \hline
 $[e^{i\hO t}\hAD e^{-i\hO t}]_{\rm n,n'}$ &
    $e^{i(\omega_{n}-\omega_{n'})t}\hAD_{\rm n,n'}$  & $e^{i(\omega_{n}+ \omega_{n'})t}\hAD_{\rm n,n'}$ &
    $e^{i(-\omega_{n}-\omega_{n'})t}\hAD_{\rm n,n'}$ & $e^{i(-\omega_{n}+ \omega_{n'})t}\hAD_{\rm n,n'}$ \\
 $\hAD_1(t)_{\rm n,n'}$ & $e^{ i(\omega_{n}-\omega_{n'})t}\hAD_{\rm n,n'}$ & 0 &$e^{i(-\omega_{n}-\omega_{n'})t}\hAD_{\rm n,n'}$ & 0 \\
 $\hAD_2(t)_{\rm n,n'}$ & 0 & $e^{ i(\omega_{n}+\omega_{n'})t}\hAD_{\rm n,n'}$ & 0 &$e^{i(-\omega_{n}+\omega_{n'})t}\hAD_{\rm n,n'}$ \\
 \hline
 $[e^{i\hO t}\hApD e^{-i\hO t}]_{\rm n',n}$ &
    $e^{i(\omega_{n'}-\omega_{n})t} \hApD_{\rm n',n}$ & $e^{i( \omega_{n'}+ \omega_{n})t}\hApD_{\rm n',n}$ &
    $e^{i(-\omega_{n'}-\omega_{n})t}\hApD_{\rm n',n}$ & $e^{i(-\omega_{n'}+ \omega_{n})t}\hApD_{\rm n',n}$ \\
 $\hApD_1(t)_{\rm n',n}$ &$e^{i(\omega_{n'}-\omega_{n})t} \hApD_{\rm n',n}$&0 &   $e^{i( \omega_{n'}+ \omega_{n})t}\hApD_{\rm n',n}$&0 \\
 $\hApD_2(t)_{\rm n',n}$ &0 &$e^{i(-\omega_{n'}-\omega_{n})t}\hApD_{\rm n',n}$&0 &$e^{i(-\omega_{n'}+ \omega_{n})t}\hApD_{\rm n',n}$   \\
 \hline
\end{tabular}
\caption{Three upper rows: matrix elements of the Heisenberg operator $\hAD(t)=e^{i\hO t}\hAD(0)e^{-i\hO t}$ and
         matrix elements of the explicit form of $\hAD(t)=\hAD_1(t)+\hAD_2(t)$, as given in
         Eqs.~(\ref{H_Init_A1}) and~(\ref{H_Init_A2}), calculated for four
         combinations of $(\epsilon,\epsilon')$. Three lower rows: the same for the operator
         $\hApD(t)=e^{i\hO t}\hAD(0)e^{-i\hO t}$ and the explicit form $\hApD(t)=\hApD_1(t)+\hApD_2(t)$.}
\end{table}

\end{widetext}

We want to prove equivalence of the general Heisenberg form of operators $\hAD(t)=e^{i\hO t}\hAD(0)e^{-i\hO t}$
and their explicit time-dependent form given in Eqs.~(\ref{H_Init_A1}) and~(\ref{H_Init_A2}).
We do this by showing that the matrix elements of~$\hAD(t)$ obtained by the Heisenberg
formula and by using Eqs.~(\ref{H_Init_A1}) and~(\ref{H_Init_A2}) are the same.
To calculate the matrix elements we take
two eigenstates of the operator~$\hO$: $|{\rm n}\rangle=|n,k_x,k_z,\epsilon,s\rangle$ and
$|{\rm n'}\rangle=|n',k_x',k_z',\epsilon',s' \rangle$ with~$n'=n+1$.
We use Eq.~(\ref{M_AD1nnp}) for the matrix element of~$\hAD_1(t)$ and Eq.~(\ref{M_AD2nnp})
for the matrix element of~$\hAD_2(t)$. On the other hand, we calculate the
matrix elements of $e^{i\hO t}\hAD(0)e^{-i\hO t}$.
We compare the matrix elements calculated by the two methods for all
combinations of the band indexes $\epsilon, \epsilon'$.
Writing $\omega_n = E_{n,k_z}/\hbar$,
$\omega_{n'}= E_{n',k_z}/\hbar$, and $\lambda_{n,k_z}=\omega_{n'}$ we obtain
results summarized in Table 1. It is seen that the matrix elements of $\hAD(t)=e^{i\hO t}\hAD(0)e^{-i\hO t}$
are equal to the matrix elements of $\hAD(t)= \hAD_1(t)+\hAD_2(t)$.
Since the states $|{\rm n}\rangle$ form a complete set, the equality holds for every matrix element of~$\hAD(t)$.
This way we proved the equivalence of the two forms of~$\hAD(t)$.
It is to be noted that selecting~$\nu=-1$ instead of~$\nu=+1$ in the definition of the square root of
operator~$\hM^2$, see Eqs.~(\ref{M_fM}), leads to the same results.

\section{} \label{AppendixUmn}
Here we consider some properties of the coefficients ~$U_{m,n}$, as defined in Eq.~(\ref{Avg_Umn}).
First, we prove the sum rule $\sum_n U_{n,n}=1$.
Let $|n,k_x\rangle$ be an eigenstate of the Hamiltonian $\hat{H}=(\hbar^2/2m)(\hat{\bm p} -e{\bm A})^2$.
In the standard notation there is
$\langle {\bm r}|n,k_x\rangle=e^{ik_xx}{\rm H}_{\rm n}(\xi)e^{-\xi^2/2}/\sqrt{L}C_n$.
For any normalized state~$|f\rangle$ we have
\begin{equation}
 1=\langle f|f\rangle = \sum_{n=0}^{\infty}\int_{-\infty}^{\infty} dk_x
    \langle f|n,k_x\rangle \langle n,k_x|f\rangle.
\end{equation}
Since $F_n(k_x)=\langle n,k_x|f\rangle$, see Eq.~(\ref{Avg_Fn}), we obtain
\begin{equation} \label{App2_1}
 1=\sum_{n=0}^{\infty}\int_{-\infty}^{\infty} F_n^*(k_x) F_n(k_x)\ dk_x = \sum_{n=0}^{\infty} U_{n,n}.
\end{equation}
This proves the normalization of $U_{n,n}$. Since the integral in Eq.~(\ref{App2_1}) can be expressed as
$\int_{-\infty}^{\infty}|F_n(k_x)|^2\ dk_x$, it is seen that~$U_{n,n}$ are non-negative.
The above sum rule was used to: i) verify the accuracy of numerical computations of~$U_{m,n}$,
ii) estimate the truncation of infinite series appearing in the calculation of~$\hY(t)$ and~$\hX(t)$.

Now we calculate another sum rule.
Consider an average value~$J$ of the operator~$\hap$ over a two-dimensional
wave packet $J=\langle f_{xy}|\hap|f_{xy}\rangle$. Inserting the unity
operator $1=\sum_n\int dk_x|n,k_x\rangle\langle n,k_x|$ we have
\begin{equation}
J=\langle f_{xy}|\hap|f_{xy}\rangle =
 \sum_{n=0}^{\infty}\int_{-\infty}^{\infty} dk_x\langle f_{xy}\hap|n,k_x\rangle \langle n,k_x|f_{xy}\rangle.
\end{equation}
Using the definitions of~$F_n(k_x)$ and~$U_{m,n}$ [see Eqs.~(\ref{Avg_Fn}) and~(\ref{Avg_Umn})] we obtain
\begin{eqnarray}
J&=&\sum_{n=0}^{\infty}\int_{-\infty}^{\infty}
 \langle f_{xy}|n+1,k_x\rangle \langle n,k_x|f_{xy}\rangle \sqrt{n+1}\ dk_x
 \nonumber \\
 &=&\sum_{n=0}^{\infty} \int_{-\infty}^{\infty} \sqrt{n+1}\ F_{n+1}^*(k_x) F_{n}(k_x)\ dk_x
 \nonumber \\
 &=& \sum_{n=0}^{\infty}\sqrt{n+1}\ U_{n+1,n}.
\end{eqnarray}
To calculate~$J$ independently we take the wave packet
\begin{equation}
f_{xy}(x,y) = \frac{1}{\sqrt{\pi d_x dy}} \exp\left(-\frac{x^2}{2d_x^2} -\frac{y^2}{2d_y^2}+ ik_{0x}x \right),
\end{equation}
and calculate~$J$ inserting the unity operator $1=\int dk_x|k_x\rangle\langle k_x|$. This gives
\begin{eqnarray}
J&=&\int_{-\infty}^{\infty} \langle f_{xy}|k_x\rangle \hap \langle k_x|f_{xy}\rangle\ dk_x dy
 \nonumber \\
 &=&\int_{-\infty}^{\infty}\hspace{-0.75em}
   g_{xy}^*(k_x,y)\frac{1}{\sqrt{2}}\left(\xi-\frac{\partial}{\partial \xi}\right)
   g_{xy}(k_x,y)\ dk_x dy. \ \ \ \ \ \
\end{eqnarray}
Since $\xi=y/L-k_xL$, and $\partial/\partial \xi = L\partial/\partial y$,
the integrations over~$d_y$ and~$k_x$ are elementary and we find
\begin{equation} J=\sum_{n=0}^{\infty}\sqrt{n+1}\ U_{n+1,n} = -\frac{k_{0x}L}{\sqrt{2}}. \end{equation}
The above sum rule was used for an additional verification of~$U_{n+1,n}$ terms and
for the analytical calculation of motion of a non-relativistic electron,
see Eqs.~(\ref{LFL_y}) and~(\ref{LFL_x}).

\section{} \label{AppendixV}
Here we calculate the average electron velocity, limiting our discussion to a packet with
the second nonzero component. The~$x$ and~$y$ components of the velocity
are the time derivatives of $\langle \hX(t)\rangle^{2,2}$ and $\langle \hY(t)\rangle^{2,2}$.
Since $\langle \hX(t)\rangle^{2,2}$ and $\langle \hY(t)\rangle^{2,2}$ are combinations
of $\langle \hAD(t)\rangle^{2,2}$ and $\langle \hApD(t)\rangle^{2,2}$ [see Eqs.~(\ref{H_Yt}) and~(\ref{H_Xt})]
we calculate the time derivatives of $\langle \hAD(t)\rangle$ and $\langle \hApD(t)\rangle$,
as given in Eqs.~(\ref{Avg_22_At}) and~(\ref{Avg_22_Apt}), respectively. The average velocities are
\begin{eqnarray} \label{V_y}
\langle v_y(t)\rangle^{2,2} &=& \frac{L}{2\sqrt{2}} \sum_n \sqrt{n+1}\ \left(U_{n,n+1}+U_{n+1,n}\right)
 \times \nonumber \\
  &&\left(\frac{\partial I^+_c}{\partial t} + \frac{\partial I^-_c}{\partial t} \right),
 \\ \label{V_x}
\langle v_x(t)\rangle^{2,2} &=& \frac{L}{2\sqrt{2}} \sum_n \sqrt{n+1}\ \left(U_{n,n+1}+U_{n+1,n}\right)
 \times \nonumber \\
  &&\left(\frac{\partial I^+_s }{\partial t}+ \frac{\partial I^-_s}{\partial t} \right), \ \ \ \ \
\end{eqnarray}
where
\begin{eqnarray} \label{V_Ic}
 L\frac{\partial I^{\pm}_c}{\partial t} &=&
   \pm \sqrt{2}c \int_{-\infty}^{\infty}\frac{\hbar\omega}{E_{n+1,k_z}}|g_z(k_z)|^2 \times \nonumber \\
           && \ \ \ \ \ \ \sin\left[(E_{n+1,k_z} \mp E_{n,k_z})t/\hbar \right]dk_z,\\
                 \label{V_Is}
 L\frac{\partial I^{\pm}_s}{\partial t} &=&
 \mp \sqrt{2}c \int_{-\infty}^{\infty} \frac{mc^2 \hbar\omega}{E_{n,k_z}E_{n+1,k_z}}
           |g_z(k_z)|^2 \times \nonumber \\
           && \ \ \ \ \ \ \cos\left[(E_{n+1,k_z} \mp E_{n,k_z})t/\hbar \right]dk_z.
\end{eqnarray}
In the above equations we used $E_{n+1,k_z}^2-E_{n,k_z}^2=\hbar^2\omega^2$.
It is seen from Eqs.~(\ref{V_Ic}) and~(\ref{V_Is}) that the
integrals $(L\partial I^{-}_c/\partial t)$, describing the cyclotron motion, and
the integrals $(L\partial I^{+}_c/\partial t)$, corresponding to the ZB motion,
have {\it the same} factor $(\hbar \omega/E_{n,k_z})|g_z(k_z)|^2$.
Integrals $(L\partial I^{-}_s/\partial t)$ and $(L\partial I^{+}_s/\partial t)$
have the same property. Therefore the amplitudes of the cyclotron velocity and the ZB velocity
are of the same order of magnitude. On the other hand, the amplitudes of positions differ by several
orders of magnitude.

Alternatively, we calculate the average velocities for the canonical velocity operators.
The velocity operator is obtained from the equation of motion
$\hat{\bm v} = (i/\hbar)[\hHD,\hat{\bm r}]$, which gives $\hat{v}_x=c\hat{\alpha}_x$
and $\hat{v}_y=c\hat{\alpha}_y$.
Now we show that the average velocities obtained in Eqs.~(\ref{V_y})-(\ref{V_x})
are equal to the averages of $\hat{v}_y(t)$ and $\hat{v}_x(t)$. We limit our calculations to a wave packet
with the second non-zero component.

The average of $\hat{v}_x(t)=e^{i\hHD t/\hbar} (c\hat{\alpha}_x) e^{-i\hHD t/\hbar}$ is
\begin{equation}
\langle \hat{v}_x(t)\rangle^{2,2} = c\sum_{\rm n,n'}\langle f|{\rm n}\rangle(\hat{\alpha}_x)_{\rm n,n'}
    \langle {\rm n'}|f\rangle e^{i(E_{\rm n}-E_{\rm n'})t/\hbar}.
\end{equation}
From Eq.~(\ref{Avg_Pak_nf}) we have $\langle {\rm n}|f\rangle = \chi_{n\epsilon k_z} g_z(k_z)s_2F_{n}(k_x)$ and
the matrix element $(\hat{\alpha}_x)_{\rm n,n'}$ is straightforward.
The summation in $\langle \hat{v}_x(t)\rangle^{2,2}$
over $s_1$ and $s_2$ gives two non-vanishing terms. We have
\begin{eqnarray}
\langle \hat{v}_x(t)\rangle^{2,2} &=& -c\sum_{n,n',\epsilon,\epsilon'} \int_{-\infty}^{\infty}
       dk_x dk_z \chi_{n\epsilon k_z}^2
       N_{n'\epsilon' k_z}\chi_{n'\epsilon' k_z} \times \nonumber \\
     &&  \hbar\omega_{n'}  e^{i(\epsilon E_{n,k_z}-\epsilon'E_{n',k_z})t/\hbar} \times \nonumber \\
     && \left(\delta_{n',n+1}+ \delta_{n',n-1}   \right) |g_z(k_z)|^2.
\end{eqnarray}
There is $\chi_{n\epsilon k_z}^2 =(1+\epsilon mc^2)/(2E_{n,k_z})$ and
$N_{n\epsilon k_z}\chi_{n\epsilon k_z} = \epsilon/(2E_{n,k_z})$. Performing the summation over $n'$,
integration over $k_x$ and replacing in the second term $n\rightarrow n+1$ we obtain
\begin{eqnarray} \label{V_alphax}
\langle \hat{v}_x(t)\rangle^{2,2} &=&
      -\frac{c}{4}\sum_{n,\epsilon,\epsilon'}\sqrt{n+1}\ U_{n,n+1}
       \int_{-\infty}^{\infty} dk_z |g_z(k_z)|^2 \times \nonumber  \\
     &&\left(1+\frac{\epsilon mc^2}{E_{n,k_z}}\right)\frac{\epsilon' \hbar\omega}{E_{n+1,k_z}}
       e^{i(\epsilon E_{n,k_z}-\epsilon'E_{n+1,k_z})t/\hbar} \nonumber \\
     &-&\frac{c}{4}\sum_{n,\epsilon,\epsilon'}\sqrt{n+1}\ U_{n+1,n}
       \int_{-\infty}^{\infty} dk_z |g_z(k_z)|^2 \times \nonumber  \\
     &&\left(1+\frac{\epsilon' mc^2}{E_{n,k_z}}\right)\frac{\epsilon \hbar\omega}{E_{n+1,k_z}}
       e^{i(\epsilon E_{n+1,k_z}-\epsilon'E_{n,k_z})t/\hbar}. \nonumber \\
\end{eqnarray}
There is
\begin{eqnarray} \label{V_cosine}
   \frac{1}{4}\sum_{\epsilon,\epsilon'} \epsilon\epsilon'
  e^{i(\epsilon E_{n} - \epsilon E_{n+1})t/\hbar} = \nonumber  \\
    \cos\left[\frac{(E_{n+1}-E_{n})t}{\hbar}\right]-  \cos\left[\frac{(E_{n+1}+E_{n})t}{\hbar}\right],
\end{eqnarray}
and the summations over the two terms with single $\epsilon$ and $\epsilon'$ cancel out.
Rearranging terms in Eq.~(\ref{V_alphax})
we obtain the same result for  $\langle \hat{v}_x(t)\rangle^{2,2}$ as in Eq.~(\ref{V_x}).
Calculations for $\langle \hat{v}_y(t)\rangle^{2,2}$ are similar to those given above.
Since $\hat{\alpha}_y$ has both positive and negative anti-diagonal elements, the expression for
$\langle \hat{v}_y(t)\rangle^{2,2}$  in Eq.~(\ref{V_alphax}) has two terms with opposite signs. Therefore the
summation over $\epsilon,\epsilon'$ cancels out the terms containing cosine functions, which appear
in Eq.~(\ref{V_cosine}), and only terms with sine function survive. After rearranging these terms we
also recover Eq.~(\ref{V_y}). This way we showed that the average velocity obtained from the differentiation
of $\langle \hat{y}(t)\rangle^{2,2}$ and  $\langle \hat{x}(t)\rangle^{2,2}$ are equal to the
average values of operators $\langle c\hat{\alpha}_y(t)\rangle^{2,2}$ and  $\langle c\hat{\alpha}_x(t)\rangle^{2,2}$.

\section{}  \label{Appendix4Packets}

\begin{figure}
\includegraphics[width=8.5cm,height=8.5cm]{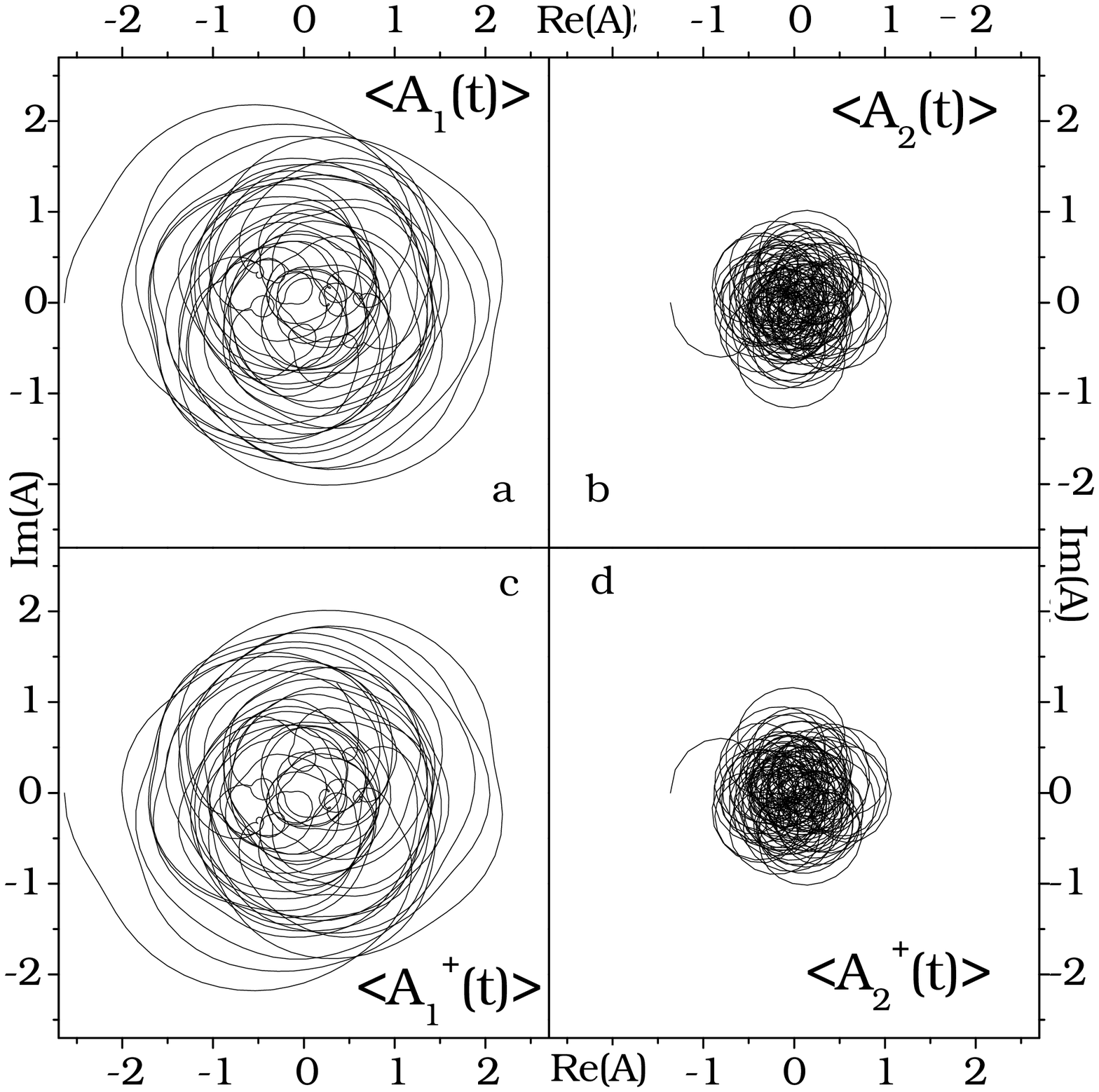}
\caption{Calculated time evolution of dynamic averages:
         a) $\langle \hA_1(t)\rangle^{2,2}$,
         b) $\langle \hA_2(t)\rangle^{2,2}$,
         c) $\langle \hAp_1(t)\rangle^{2,2}$,
         d) $\langle \hAp_2(t)\rangle^{2,2}$,
         as given in Eqs.~(\ref{App_4p1})-(\ref{App_4p4}), for 2+1 DE. Trap parameters as in
         Fig.~\ref{FigSim1}, simulated gap frequency $\Omega=2\pi \times 4000$ Hz.
         Packet parameters: $d_x=0.63\lambda_c$, $d_y=0.57\lambda_c$, $k_{0x}=0.999\lambda_c^{-1}$.
         Motion is plotted for $0 < t < 8 $ ms.} \label{FigA1}
\end{figure}

Below we analyze the structure of electron motion.
Time evolution of the average values of~$\hAD(t)$ and~$\hApD(t)$ is equivalent to
the evolution of {\it four } sub-packets: $\langle \hAD_1(t)\rangle$, $\langle \hAD_2(t)\rangle$,
$\langle \hApD_1(t)\rangle$, $\langle \hApD_2(t)\rangle$, see Eqs.~(\ref{Avg_22_At})-(\ref{Avg_22_Apt}).
We take the packet $\langle {\bm r} | f\rangle = (0,f({\bm r}), 0,0)^T$ and follow the method
similar to that presented in the calculation of $\langle \hAD_1\rangle$ in Eq.~(\ref{Avg_22_A_0}).
For simplicity we consider the 2+1 Dirac equation setting $|g_z(k_z)|^2\rightarrow \delta(k_z)$, which gives
\begin{eqnarray}
\lefteqn{\langle \hAD_1(t)\rangle^{2,2} =} \nonumber \\
&&\sum_n \sqrt{n+1}\ U_{n,n+1} \sum_{\epsilon,\epsilon'}
e^{i(\epsilon E_{n,0}-E_{n+1,0})t/\hbar}
\nonumber \\
&& \times \frac{1+\epsilon'}{4}\left[1 + \epsilon\epsilon'\frac{E_{n,0}}{E_{n+1,0}}
        + mc^2\left(\frac{\epsilon}{E_{n,0}} + \frac{\epsilon'}{E_{n+1,0}}\right) \right]. \ \ \ \ \ \
\end{eqnarray}
Performing the summation over $\epsilon, \epsilon'$, and
writing $E_n=E_{n,0}$, $\omega_n^c=(E_{n+1}-E_n)/\hbar$, $\omega_n^Z=(E_{n+1}+E_n)/\hbar$,
${\cal U}_n=\sqrt{n+1}\ U_{n,n+1}$, and ${\cal U}_n^{\dagger}=\sqrt{n+1}\ U_{n+1,n}$ we obtain
\begin{widetext}
\begin{eqnarray}
\langle\hAD_1(t)\rangle^{2,2}=\frac{1}{4}\sum_n {\cal U}_n \left\{              \label{App_4p1}
   T^{++}_{++}\cos\left(\omega_n^ct\right) + T^{+-}_{+-}\cos\left(\omega_n^Zt\right)
 -iT^{++}_{++}\sin\left(\omega_n^ct\right) +iT^{-+}_{-+}\sin\left(\omega_n^Zt\right) \right\},  \\
\langle \hAD_2(t)\rangle^{2,2}=\frac{1}{4}\sum_n {\cal U}_n \left\{             \label{App_4p2}
   T^{+-}_{-+}\cos\left(\omega_n^ct\right) + T^{++}_{--}\cos\left(\omega_n^Zt\right)
 +iT^{+-}_{-+}\sin\left(\omega_n^ct\right) +iT^{++}_{--}\sin\left(\omega_n^Zt\right) \right\},  \\
\langle \hApD_1(t)\rangle^{2,2}= \frac{1}{4}\sum_n {\cal U}_n^{\dagger} \left\{ \label{App_4p3}
   T^{++}_{++}\cos\left(\omega_n^ct\right) + T^{+-}_{+-}\cos\left(\omega_n^Zt\right)
 +iT^{++}_{++}\sin\left(\omega_n^ct\right) +iT^{+-}_{+-}\sin\left(\omega_n^Zt\right) \right\},  \\
\langle \hApD_2(t)\rangle^{2,2}= \frac{1}{4}\sum_n {\cal U}_n^{\dagger} \left\{ \label{App_4p4}
   T^{+-}_{+-}\cos\left(\omega_n^ct\right) + T^{++}_{--}\cos\left(\omega_n^Zt\right)
 +iT^{-+}_{+-}\sin\left(\omega_n^ct\right) +iT^{--}_{++}\sin\left(\omega_n^Zt\right) \right\},
\end{eqnarray}
\end{widetext}
where we used the notation
\begin{equation}
 T^{s_1s_2}_{s_3s_4}= s_1 + s_2\frac{mc^2}{E_n} + s_3\frac{mc^2}{E_{n+1}} + s_4\frac{E_n}{E_{n+1}},
\end{equation}
with $s_1, s_2, s_3, s_4 = \pm 1$.
Each of the terms in Eqs.~(\ref{App_4p1})-(\ref{App_4p4}) contains sine and cosine functions with
the cyclotron and ZB frequencies. The structure of these terms is significantly different.
To see this we consider the non-relativistic limit: $E_{n+1}\simeq E_n\simeq mc^2$.
Then the motion of sub-packets $\langle \hAD_1(t)\rangle^{2,2}$ and $\langle \hApD_1(t)\rangle^{2,2}$
reduces to the cyclotron motion, while the
averages $\langle \hAD_2(t)\rangle^{2,2}$ and $\langle \hApD_2(t)\rangle^{2,2}$
vanish. The above sub-packets describe natural components
of the electron motion in a magnetic field. The direct averaging
of $\langle \hAD(t)\rangle$ or $\langle \hApD(t)\rangle$, as presented in the previous sections,
allows us to calculate the evolution
of the physical quantities but it does not exhibit the structure of the motion.
The exact operator results, as given in Eqs.~(\ref{H_Init_A1})-(\ref{H_Init_Ap2}),
provide a deeper understanding of this structure.

In Fig.~\ref{FigA1} we plot time evolutions of the four sub-packets
$\langle \hAD_1(t)\rangle^{2,2}$,
$\langle \hAD_2(t)\rangle^{2,2}$,
$\langle \hApD_1(t)\rangle^{2,2}$ and
$\langle \hApD_2(t)\rangle^{2,2}$,
calculated with the use of Eqs.~(\ref{App_4p1})-(\ref{App_4p4})
for simulated gap frequency $\Omega=2\pi \times 4000$ Hz.
At low magnetic fields, the components $\langle \hAD_2(t)\rangle^{2,2}$
and $\langle \hApD_2(t)\rangle^{2,2}$ are much smaller than $\langle \hAD_1(t)\rangle^{2,2}$
and $\langle \hApD_2(t)\rangle^{2,2}$.
Note that $\langle \hAD_1(t)\rangle^{2,2}$ spins in the opposite direction
to $\langle \hApD_1(t)\rangle^{2,2}$,
and similarly for $\langle \hAD_2(t)\rangle^{2,2}$ and $\langle \hApD_2(t)\rangle^{2,2}$.
The four components of motion are persistent for the 2+1 Dirac equation.

\section{} \label{AppendixBarut}

In this Appendix we discuss the relation of our work to that of Barut and Thacker
(BT, Ref.~\cite{Barut1985}) concerned with the same subject.
Barut and Thacker calculated the ZB of relativistic electrons in the presence of a magnetic
field at the operator level. Their work was the first treatment of this subject but, in our opinion,
it suffered from a few deficiencies.

Barut and Thacker considered the time dependence of electron motion introducing from the
beginning its~$\hat{x}$ and~$\hat{y}$ components
[in our notation, cf. Eqs.~(\ref{H_Y}) and~(\ref{H_X}) and Appendix \ref{AppendixXY}]
rather than~$\hAD$ and~$\hApD$ operators.
This choice was unfortunate since~$\hAD$ and~$\hApD$ satisfy separately important
operator equations~(\ref{H_Btt}) and~(\ref{H_Bptt}), in which
$\hB = \exp(-i\hO t)\hAD$ and $\hBp = \hApD\exp(+i\hO t)$ operators stand at the RHS and the LHS,
respectively. The operators~$\hat{x}$ and~$\hat{y}$ do not satisfy such equations and,
'forcing'~$\hat{x}$ and~$\hat{y}$ to satisfy the corresponding relations, BT introduced the frequency
$\omega_2=\sqrt{2(mc^2)^2-(\hbar\omega)^2}$ (in our notation).
The problem here is that for~$\hbar\omega >\sqrt{2}mc^2$
this frequency becomes imaginary leading to solutions growing
exponentially in time. In our treatment no such problem occurs since all the frequencies are of the
form $\omega_n=(E_{n+1,k_z}\pm E_{n,k_z})/\hbar$, i.e. they are real for all magnetic fields.

The calculation of BT gave only two interband ZB frequencies and two intraband (cyclotron resonance)
frequencies contributing to the electron motion. On the other hand, we obtain {\it two series}
of intraband and interband frequencies because the Gaussian wave packet,
which we use for the averaging procedure, includes numerous Landau eigenstates in a magnetic field.
On the other hand, BT did not introduce a wave packet projecting their operator results on the ground
electron state. In contrast to our approach the procedure of Barut and Thacker uses the proper time
formalism.

\end{document}